\documentclass[aps,prb,twocolumn,floatfix,showpacs]{revtex4-1}
\usepackage{amsmath,amssymb,graphicx,bm}
\usepackage[compat=1.1.0]{tikz-feynman}
\def\veck{\mathbf k}
\def\vecq{\mathbf q}

\def\be{\begin{equation}}
\def\ee{\end{equation}}

\begin{document}
\title{Curie-Weiss susceptibility in strongly correlated electron systems}

\author{V\'aclav  Jani\v{s} } \author{Anton\'\i n Kl\'\i\v c} \author{Jiawei Yan} \author{Vladislav Pokorn\'y}

\affiliation{Institute of Physics, The Czech Academy of Sciences, Na Slovance 2, CZ-18221 Praha  8,  Czech Republic}

\email{janis@fzu.cz}

\date{\today}


\begin{abstract}
The genesis of the Curie-Weiss magnetic response observed in most transition metals that are Fermi liquids at low temperatures has been an enigma for decades and has not yet been fully explained from microscopic principles. We show on the single-impurity Anderson model how the quantum dynamics of strong electron correlations leads  to the Curie-Weiss magnetic susceptibility sufficiently above the Kondo temperature. Such a behavior has not yet been demonstrated and  can be observed only when the bare interaction is substantially screened (renormalized) and a balance between quantum and thermal fluctuations is kept. We set quantitative criteria for the existence of the Curie-Weiss law.   
\end{abstract}

\maketitle 

\section{Introduction}

The behavior of electrons is decisive for shaping the low-temperature properties of metals. Unless in an ordered state the electrons form a Fermi liquid and display Pauli paramagnetism. This behavior seems to be independent of the strength of the electron correlations. Most of  the transition metals and their compounds seem to contradict this simple picture in that the Curie-Weiss magnetic response is observed in a broad interval of temperatures even below the Fermi temperature. 
 
The Curie law for the paramagnetic low-temperature susceptibility $\chi =  C/T$ was derived by using the concept of local magnetic moments.\cite{Langevin:1905aa} Later on, Weiss extended the Curie law to $\chi= C/(T - T_{c})$ by introducing an interaction between atomic magnetic moments in order to cover transitions to the ferromagnetic state at $T_{c}$.\cite{Weiss:1907aa} This Curie-Weiss law well reproduces the magnetic response of insulating materials with fixed spin moments and no relevant charge fluctuations.    

The metallic (itinerant) magnetism was first described by Bloch in terms of  electron waves\cite{Bloch:1929aa} to which Stoner later added a mean-field description of electron correlations.\cite{Stoner:1938aa} Such a static, weak-coupling theory with no local moments leads to Pauli paramagnetism at low temperatures and cannot account for the Curie-Weiss behavior when the critical temperature lies below the Fermi energy of the degenerate Fermi gas.\cite{Nagakawa:1957aa} 

A wave of efforts arose to understand the origin of the Curie-Weiss magnetic response in systems without the apparent presence of  local magnetic moments.\cite{Herring:1966aa}  First theory that qualitatively correctly reproduced the Curie-Weiss law in weak ferromagnetic metals was developed by Moriya and collaborators.  They introduced a self-consistent theory for the local magnetic susceptibility by including static spin fluctuations.\cite{Moriya:1973aa,Ueda:1975aa,Moriya:1978aa,Takahashi:1979aa,Moriya:1985aa,Takahashi:1986aa,Moriya:1991aa, Moriya:2006aa}  Although the theory was able to interpolate between the weak ferromagnetic and local moment pictures of itinerant magnetism, it missed the strong-coupling limit and did not provide a consistent thermodynamic  and conserving approximation. Neither did it clearly explain the microscopic origin of the Curie-Weiss behavior.  

The modern approach to strongly correlated electrons based on the dynamical mean-field theory (DMFT) was used to derive an implicit form of the Curie-Weiss law from local dynamical fluctuations\cite{Byczuk:2002aa} and in combination with the local density functional it also reproduced the critical behavior above the ferromagnetic transition of iron and nickel.\cite{Lichtenstein:2001aa}  Although DMFT suggested that local dynamical fluctuations may be responsible for the Curie-Weiss law, the microscopic mechanism behind it has not been disclosed. 

The reason for the failure of DMFT to identify the origin of the Curie-Weiss law is the lack of the two-particle renormalization, that is, a self-consistent determination of the screening of the bare interaction. The simplest systematic theory with two-particle renormalizations is the parquet construction.\cite{DeDominicis:1962aa,DeDominicis:1963aa}  It is in its full generality, however, not analytically controllable and it does not reproduce the strong-coupling Kondo limit of the single-impurity Anderson model (SIAM) correctly.\cite{Bickers:1991aa,Bickers:1991ab}        

The way out from this trap of complexity with little analytic control is to reduce the parquet scheme. One has to keep its substantial part, the two-particle self-consistency, so that to interpolate qualitatively correctly between the weak- and strong-coupling regimes in a controlled way. One of the present authors has developed such an analytically controllable scheme qualitatively correctly interpolating between the weak and strong coupling in the SIAM.\cite{Janis:2007aa,Janis:2008ab,Janis:2017aa,Janis:2017ab,Janis:2019aa} Since the parquet approach leads to Fermi liquid at low temperatures of the SIAM, it is necessary to extend this approach properly to higher temperatures and beyond the Fermi-liquid regime.  We succeeded to do so and introduced the Kondo temperature as a point at which the thermal fluctuations equal the quantum, zero-temperature ones and  above which the Fermi-liquid description breaks down.\cite{Janis:2020aa}  

The aim of this paper is to disclose the origin of the Curie-Weiss law in strongly correlated electron systems. We show that local dynamical fluctuations due to strong electron correlations 
generate the Curie-Weiss magnetic response in metallic systems. The necessary constituents of the explanation of the Curie-Weiss law in itinerant systems are  i) two-particle self-consistency renormalizing the interaction strength, ii) reliable interpolation between weak and strong coupling, and iii) balance between local quantum and thermal fluctuations.  All these conditions are met in our reduced parquet scheme.\cite{Janis:2019aa,Janis:2020aa} We apply it to the SIAM as the generic model to demonstrate how the dynamical forming of the local magnetic moment in strong coupling together with thermal fluctuations lead to a non-Fermi-liquid behavior and the Curie-Weiss susceptibility above the Kondo temperature. Our findings have the general relevance and hold for extended systems with critical magnetic fluctuations as well. We demonstrate that the Curie-Weiss law is generally caused  by local dynamical fluctuations while the spatial fluctuations affect the region of its validity.

\section{Local model of strong electron correlations }

\subsection{Model of a local Fermi liquid}

Metallic systems are distinguished by fermionic low-energy excitations of the ground state, unless strong electron correlations destroy the low-temperature Fermi-liquid regime. The strong-coupling limit of the spin and charge symmetric state of the SIAM at low temperatures (Kondo limit)  stands for a model situation of a local Fermi  liquid where the dynamical fluctuations lead to the formation of a local magnetic moment. Although the ground state remains Fermi liquid for arbitrarily strong electron correlations  a new exponentially small two-particle scale emerges in strong coupling.  The magnetic susceptibility is large, but finite as well as the effective mass of the one-particle excitations.  The spectral function displays a three-peak structure with an exponentially narrow central quasiparticle peak. The width of the central peak is proportional to the inverse lifetime of the pair of the electron with a given spin and the hole with the opposite spin. That is why the Kondo limit of the SIAM is the simplest situation where the long-lived local magnetic moment can lead to the Curie-Weiss magnetic response.

The Hamiltonian of the SIAM in second quantization is      
\begin{multline}\label{eq:H-SIAM}
  H_{I} = \sum_{{\bf k}\sigma} \epsilon({\bf k})
  c^{\dagger}_{{\bf k}\sigma} c^{\phantom{\dagger}}_{{\bf k}\sigma}+
  E_d\sum_\sigma d^{\dagger}_\sigma d_\sigma +
  U d^{\dagger}_\uparrow d_\uparrow d^{\dagger}_\downarrow  d_\downarrow
  \\ 
  +\ \sum_{{\bf
      k}\sigma}\left(V^{\phantom{*}}_{{\bf k}}d^{\dagger}_\sigma
    c^{\phantom{\dagger}}_{{\bf k}\sigma} + V^*_{{\bf k}}
    c^{\dagger}_{{\bf k}\sigma} d^{\phantom{\dagger}}_\sigma\right) \,, 
\end{multline}
where $ c^{\dagger}_{\veck \sigma}, c^{\phantom{\dagger}}_{\veck \sigma}$ are creation and annihilation operators of the conduction electrons with spin $\sigma$ and momentum $\veck$  and $d^{\dagger}_{\sigma},d_{\sigma}$ are the creation and annihilation operators of the impurity electron with spin $\sigma$. 
 
The conduction electrons can be projected out, which leads to a  band of energy states on the impurity. The effect of the conduction electrons can be approximated by a shift $\Delta= 2\pi V^{2}\rho_{c}$ of the imaginary part of the frequency of the bare propagator, where $\rho_{c}$ is the local density of states of the conduction electrons at the Fermi energy.  

The Curie-Weiss response must be deduced from the two-particle vertex. One hence has to directly approach the two-particle response and vertex functions when looking for a specific behavior of the magnetic susceptibility.  We use the standard diagrammatic perturbation theory for the full two-particle vertex that in the local model can be decomposed into the irreducible  $\Lambda$ and the reducible $\mathcal{K}$ vertices 
\begin{multline}\label{eq:Gamma-sum}
\Gamma(i\omega_{n},i\omega_{n'}; \nu_{m}) 
= \Lambda(i\omega_{n}, i\omega_{ n'};i\nu_{m}) 
\\
+\ \mathcal{K}(i\omega_{n}, i\omega_{n'}; i\nu_{m })  \,,
\end{multline}
of a specific two-particle scattering channel. Here $\omega_{n}=(2n + 1)\pi k_{B}T$ and $\nu_{m}= 2m\pi k_{B}T$ are fermionic and bosonic Matsubara frequencies, respectively. The scattering channel should be chosen so that the respective Bethe-Salpeter equation is expected to generate a divergence due to multiple scatterings of pairs of electrons or electron-hole pairs. Since we expect that the magnetic susceptibility will be large and will approach a critical point at low temperatures, we use the decoupling of the full two-particle vertex into the reducible and irreducible components in the singlet electron-hole channel with non-singular $\Lambda$  and possibly singular $\mathcal{K}$ as observed in the weak-coupling perturbation expansion.

\subsection{Reduced parquet equations}

\begin{figure}
\includegraphics[width=9cm]{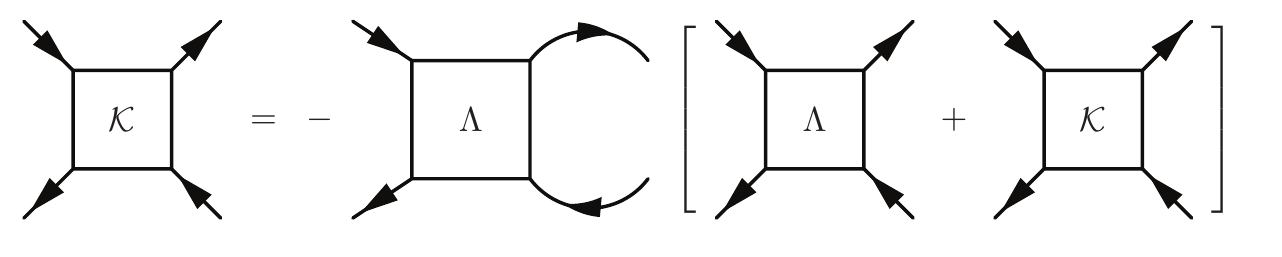}
\caption{The  reducible vertex $\mathcal{K}$  determined from the Bethe-Salpeter equation with the integral kernel $\Lambda$, being the vertex  irreducible  with respect to multiple scatterings of the singlet electron-hole pairs. The arrows indicate the charge propagation and the upper lines carry spin up while the lower lines spin down.     \label{fig:RPE-eh} }
\end{figure}

The problem of the weak-coupling diagrammatic (perturbation) expansion is that it will reach a singularity in the two-particle vertex in intermediate coupling. We hence need to go over to non-perturbative approximations if we want to describe reliably the transition from weak-coupling to strong-coupling regimes. Multiple scatterings of the singlet electron-hole pairs lead to a singularity in the Bethe-Salpeter equation with the bare repulsive interaction, the pole in the two-particle vertex from the random-phase approximation (RPA). The corresponding Bethe-Salpeter equation for the reducible vertex $\mathcal{K}$ in the electron-hole singlet channel is graphically represented in Fig.~\ref{fig:RPE-eh}.  If the full irreducible vertex $\Lambda$ is replaced with the bare interaction $U$ the reducible vertex $\mathcal{K}$  becomes singular at the critical interaction $U_{c} = \pi \Delta$. This singularity is, however, unphysical in the local models with no spatial fluctuations. We have to renormalize the bare interaction to a self-consistent equation for the irreducible vertex $\Lambda$.  It will be achieved via a two-particle self-consistency of the parquet equations. 

The complete set of the parquet equations contains three coupled two-particle scattering channels and their solution can be reached only numerically in the Matsubara formalism.\cite{Bickers:1991ab,Yang:2009aa,Li:2016aa,Li:2019aa,Eckhardt:2020aa} The electron-hole singlet and triplet channels generate the same singularity in the two-particle vertex. We can hence neglect one of them without qualitatively affecting the critical behavior of the two-particle vertex. This is why we can resort only to two-channel parquet equations. The full solution of neither the three channel nor the two-channel parquet equations leads to the Kondo behavior.\cite{Janis:2006ab} The parquet equations with the full one-particle self-consistency also fail to guarantee that the strength of the electron repulsion is determined only by the present charge densities.\cite{Janis:1998aa,Janis:2017aa} 

\begin{figure}
\includegraphics[width=8cm]{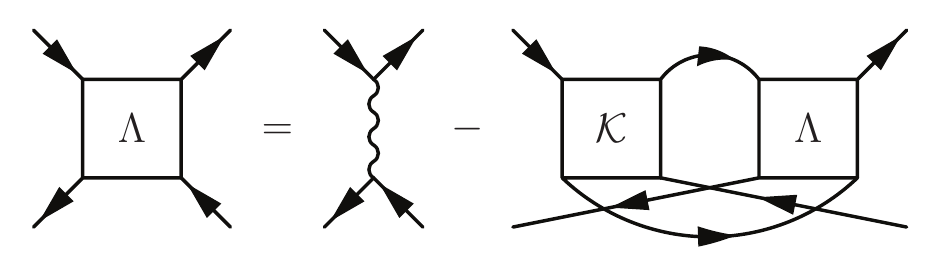}
\caption{The irreducible vertex $\Lambda$, a renormalized interaction strength,  determined from a reduced Bethe-Salpeter equation with multiple electron-electron scatterings in which only the dominant contributions from the reducible vertex $\mathcal{K}$ that do not drive the solution out of its critical region are taken into account as explained in the text.  The vertical wavy line is the bare Hubbard interaction between electrons with opposite spins.  \label{fig:RPE-ee} }
\end{figure}

The irreducible vertex $\Lambda$ renormalizing the bare interaction $U$ from the RPA in the two-channel parquet scheme  is determined  from the Bethe-Salpeter equation with multiple scatterings of the singlet electron pairs. The renormalization obtained from the full Bethe-Salpeter equation misses the strong-coupling Kondo regime in the SIAM. We hence introduced a reduced Bethe-Salpeter equation in the electron-electron channel that renormalizes the interaction appropriately so that to recover qualitatively correctly the Kondo strong-coupling asymptotics.\cite{Janis:2019aa} The diagrammatic representation of this equation is plotted in Fig.~\ref{fig:RPE-ee}.  

The reasoning behind its derivation is as follows.  The bare interaction is used in the weak-coupling regime until the critical region of the RPA pole is reached. A new small two-particle scale emerges. We call it the Kondo scale and denote its dimensionless form $a$. We now rearrange the perturbation expansion according to the powers of the inverse Kondo scale $a^{-1}$. The leading term in the irreducible vertex $\Lambda^{ee}$ in the Bethe-Salpeter equation with multiple scatterings of the singlet electron pairs is the reducible vertex from the electron-hole channel and  we replace $\Lambda^{ee}\to\mathcal{K}$. If we resort to the parquet equations with the bare interaction as the fully irreducible vertex we must renormalize only the irreducible vertex $\Lambda$ by multiple scatterings of the singlet electron pairs and suppress the self-renormalization of the singular vertex $\mathcal{K}$.  The renormalization of vertex $\mathcal{K}$ drives the solution away from the critical region of the RPA pole. The full vertex $\Gamma$ on the right-hand side of the full Bethe-Salpeter equation in the particle channel is then replaced by the irreducible one,  $\Gamma\to\Lambda$.  In this way the two-particle self-consistency of the parquet approach is  conserved and the critical behavior of the RPA pole is thereby transferred to the strong coupling where it goes over into the Kondo critical regime.  The qualitative behavior of the Kondo limit of the SIAM, that is, the linear dependence of the Kondo temperature on the bare interaction strength with the critical point at $U=\infty$ is thereby reproduced.\cite{Janis:2017aa,Janis:2019aa}  The reduced parquet equations are justified in the critical region of the reducible vertex $\mathcal{K}$  with a small scale $a\ll 1$ measuring the distance to the critical point with $a=0$. Their solution can be, however, extended outside the critical region   simulating well the qualitative behavior of the model in the whole range of the input parameters. The reduced parquet equations are the simplest approximation reproducing qualitatively correctly the quantum critical behavior due to singularities in the Bethe-Salpeter equations.  

\subsubsection{Mean-value approximation}

The vertices from the reduced parquet equations remain frequency dependent. The reduced parquet equations from Fig.~\ref{fig:RPE-eh} and Fig.~\ref{fig:RPE-ee} contain frequency convolutions.  They can hence be solved with unrestricted frequency dependence of the vertices only numerically in the Matsubara formalism. We are predominantly interested in the critical behavior of the reducible vertex $\mathcal{K}$  with a small Kondo scale $a\ll 1$ where we can separate the critical and non-critical dynamical fluctuations. The former fluctuations scale critically with the controlling parameter $a$ while the latter not. We then neglect the finite noncritical fluctuations, represented by fermionic Matsubara frequencies, and keep only the dominant critical ones that have bosonic character and drive  the reducible vertex $\mathcal{K}$ towards its critical point with $a=0$. The convolutions in the fermionic frequencies are then simplified in the spirit of the mean-value theorem so that only the asymptotic limit of the fermionic frequencies to the Fermi energy survives while the whole spectrum of the transfer bosonic frequencies of the two-particle excitations is considered without restrictions.   We thereby simplify the reduced parquet equations to a single mean-field-like self-consistent equation  for the static irreducible vertex in the electron-hole channel $\Lambda$, an effective screened interaction\cite{Janis:2019aa}  
\begin{equation}\label{eq:Lambda-static}
\Lambda 
 =  \frac U{1 - \Lambda^{2}\phi(0) X} \,,
\end{equation}
where the frequency-dependent electron-hole bubble is
\begin{multline}\label{eq:phi-Omega}
\phi(\omega_{+})  = -\int_{-\infty}^{\infty} \frac{dx}{\pi} f(x)  \left[G(x + \omega_{+}) + G(x - \omega_{+})  \right] 
\\
\times \Im G(x_{+}) \,
\end{multline}
and $\omega_{+} = \omega + i0^{+}$ denotes the way the real axis is reached from the complex plane.  The electron-electron multiple scatterings  from vertex $\mathcal{K}$ contribute to the screening of the interaction strength via  integral $X= X_{0} + \Delta X$ that we decompose into  quantum and thermal contributions, $X_{0}$ and $\Delta X$, respectively,   
\begin{subequations}\label{eq:IntegralX}
\begin{equation}\label{eq:IntegralX0}
X_{0} = -\int_{-\infty}^{\infty}\frac{dx}{\pi} f(x) \Im\left[\frac {G(x_{+}) G(-x_{+})}{1 + \Lambda\phi(-x_{+})} \right]\,,
\end{equation}
\begin{multline}\label{eq:IntegralDeltaX}
 \Delta X  = \int_{-\infty}^{\infty}\frac{dx}{\pi} \frac{\Re\left[ G(x_{+})G(-x_{+})\right]}{\sinh(\beta x)} 
 \\  
\times \Im\left[ \frac 1{1 + \Lambda\phi(-x_{+})}\right]  \,.
\end{multline}
\end{subequations}
We used an equality $f(x) + b(x) = 1/\sinh(\beta x)$. We straightforwardly continued analytically the sums over the Matsubara frequencies to spectral integrals with Fermi, $f(x) = 1/(e^{\beta x} + 1)$ and Bose, $b(x) = 1/(e^{\beta x} - 1)$, distributions. We use in this paper the bare Green function of the SIAM  $G(\omega_{\pm})= 1/(\omega - E_{F} - Un/2 \pm i\Delta)$, where $n$ is the charge density. We showed that we reproduce the exact Kondo limit qualitatively correctly  with  this propagator.\cite{Janis:2017aa,Janis:2019aa} The Kondo strong-coupling criticality is manifested at half filling, that is  for $n=1$ and $E_{F} = -U/2$, to which we resort.

The resulting approximation with the renormalized interaction $\Lambda$ has a well established analytic structure. It is easily numerically solvable in the whole range of the interaction strength and arbitrary temperature. It leads to thermodynamic properties defined analogously as in the RPA. The major difference to the RPA is that vertex $\Lambda$ significantly renormalizes the bare interaction strength and never goes over the critical interaction of the RPA, $U_{c}= \pi\Delta$ in the impurity models. Its strong-coupling asymptotics at zero temperature  is  $\Lambda\doteq \pi\Delta\left[1 - \exp(-U/\pi \Delta)\right]$.\cite{Janis:2019aa}  This approximation also qualitatively correctly interpolates between the low and high temperatures.\cite{Janis:2020aa} 

The static thermodynamic susceptibility in this approximation has a simple, mean-field-like representation
\begin{align}\label{eq:chiinv-full}
\frac{\chi^{T}}{\mu_{0}\mu_{B}^{2}} &= - \frac {2\phi(0)}{1 + \phi(0)\Lambda} \,,
\end{align}
where $\mu_{0}$ is the permeability of vacuum and  $\mu_{B}$ is the Bohr magneton. We introduce a dimensionless Kondo scale $a$, the denominator of the susceptibility  that measures the distance from the critical point at which it would vanish. The equations for the Kondo scale and the effective interaction $\Lambda$ from Eq.~\eqref{eq:Lambda-static} then are
\begin{subequations}\label{eq:a-Lambda}
\begin{align}\label{eq:a-def}
a & = 1 + \phi(0)\Lambda \,,
\\  \label{eq:Lambda-m}
 \Lambda &= \frac 1{2 X(1 - a)}\left[\sqrt{1 + 4(1 - a)U X}  - 1\right] \,.
\end{align} 
\end{subequations}
We can exclude variable $a$ from the above equations and obtain an explicit solution for the dimensionless parameter $\bar{a} = 1 - a \in (0,1)$ as a root of a cubic equations
\begin{equation}\label{eq:bara-cubic}
\bar{a}^{3} + \frac{\bar{a}}{y} = \frac{u}{y} \,, 
\end{equation} 
where we introduced dimensionless parameters $u = U|\phi(0)|$ and $y= X/|\phi(0)|$. 
We transform the cubic equation~\eqref{eq:bara-cubic} to a quadratic one for  new variable  $w^{3}$ by a substitution 
\begin{subequations}\label{eq:bara-explicit}
\begin{equation}\label{eq:bara-w}
\bar{a} = w - \frac {1}{3y w} \,.
\end{equation}
The root obeying the correct boundary conditions is 
\begin{equation}\label{eq:bara-w3}
w^{3} = \frac{u}{2y}\left[1 + \sqrt{1 + \frac 4{27} \frac 1{u^{2}y}} \right] \,.
\end{equation}
\end{subequations} 
We use the solution for $\Lambda= \bar{a}/|\phi(0)|$ from Eqs.~\eqref{eq:bara-explicit} in Eqs.~\eqref{eq:IntegralX} to close a self-consistent equation for the sum of  the integrals $y = y_{0} + \Delta y$.  The equations for the two dimensionless parameters $y_{0}$ and $\Delta y$ read
\begin{subequations}\label{eq:IntegralX-bara}
\begin{equation}\label{eq:IntegralX0-bara}
y_{0} = -\int_{-\infty}^{\infty}\frac{dx}{\pi} f(x) \Im\left[\frac {G(x_{+}) G(-x_{+})}{|\phi(0)| + \bar{a}\phi(-x_{+})} \right]\,,
\end{equation}
\begin{multline}\label{eq:IntegralDeltaX-bara}
 \Delta y  = \int_{-\infty}^{\infty}\frac{dx}{\pi} \frac{\Re\left[ G(x_{+})G(-x_{+})\right]}{\sinh(\beta x)} 
 \\  
\times \Im\left[ \frac 1{|\phi(0)| + \bar{a}\phi(-x_{+})}\right]  \,.
\end{multline}
\end{subequations}
These equations can be solved iteratively in $\bar{a}$, starting with $\bar{a} = 0$ in Eqs.~\eqref{eq:IntegralX-bara}. The integrals $y_{0}$ and $\Delta y$ are then used in Eq.~\eqref{eq:bara-w3} from which we determine the new $\bar{a}$ from Eq.~\eqref{eq:bara-w} until convergence is reached.

 \subsubsection{Low-frequency asymptotics}

We need to evaluate integrals in Eqs.~\eqref{eq:phi-Omega} and~\eqref{eq:IntegralX}. We use the low-frequency approximation in calculating the electron-hole bubble $\phi(\omega_{+})$ in the Kondo regime, where the Kondo scale $a=1 + \phi(0)\Lambda\ll 1$. We replace $1 +\phi(\omega_{\pm})\Lambda  \approx a  \mp iA\omega/\Delta$.  Such an approximation is well justified in the critical region of a singularity (pole) in the Bethe-Salpeter equation in the electron-hole channel in Fig.~\ref{fig:RPE-eh}.

We denote
\begin{equation}\label{eq:g-full}
g = \phi(0) =  \int_{0}^{\infty} \frac{dx}{\pi}  \tanh\left( \frac{\beta x}2\right)\Im\left[G(x_{+})^{2} \right]
\end{equation}
and use the bare Green function $G(x_{+})= 1/(x + i\Delta)$ to evaluate the integrals in the  definitions of the parameters to be determined from the reduced parquet equations.  We showed in our previous publications that the unperturbed Green function used in the perturbation expansion gives the best and qualitatively correct estimate  for the Kondo strong-coupling asymptotics in the SIAM.\cite{Janis:2007aa,Janis:2017aa,Janis:2019aa} The results for the dimensionless integrals $y_{0}$ and $\Delta y $ from Eqs.~\eqref{eq:y-lowT} in the low-frequency approximation $1 + \Lambda\phi(\omega_{+}) \doteq a - i\omega A/\Delta$ are 
\begin{subequations}\label{eq:X}
\begin{multline}\label{eq:X0}
 y_{0}=  - \frac 1{|g|} \int_{0}^{\infty} \frac{dx}{\pi} \tanh\left( \frac{\beta x}2\right)
    \\
  \times \frac{\Re\left[G(x_{+})^{2}\right]Ax  +  \Im \left[G(x_{+})^{2}\right] a }{ a^{2} + A^{2}x^{2}/\Delta^{2}}
\end{multline}
and
\be\label{eq:DeltaX}
\Delta y = -\frac{2 A}{\pi|g|} \int_{0}^{\infty} \frac {dx\ x}{\sinh(\beta x)}\frac{\Re\left[ G(x_{+})^{2}\right]}{a^{2} + A^{2}x^{2}/\Delta^{2}}  \,.
\ee
\end{subequations}
Here we used the electron-hole symmetry, $G(x_{+}) = - G(-x_{+})$ and $\phi(x_{+}) = \phi(-x_{+})$.

The expansion parameter of this low-frequency approximation is  
\begin{equation}\label{eq:Dprime}
A =\frac{(1 - a)\beta \Delta}{2\pi|g|}\int_{0}^{\infty}\frac{dx }{\cosh^{2}(\beta x/2)} |\Im G(x_{+})|^{2} \,.
\end{equation}
These integrals are the input for the self-consistent Eqs.~\eqref{eq:bara-explicit}.

\section{Temperature behavior}

The approximation defined by Eqs.~\eqref{eq:bara-explicit} and~\eqref{eq:IntegralX-bara} leads to a Fermi liquid at low temperatures. The Curie-Weiss susceptibility may hence be observed only at higher temperatures. The full numerical solution of these equations at non-zero temperatures is unsuitable for identifying the microscopic origin of the Curie-Weiss magnetic susceptibility. The integrals to be evaluated contain trigonometric functions preventing us from obtaining explicit analytic results. To reach analytic estimates one has to approximate the integrals with the Fermi and Bose distribution functions.

\subsection{Approximate spectral integrals}

\begin{figure*}\hspace*{-8mm}
\includegraphics[width=7.5cm]{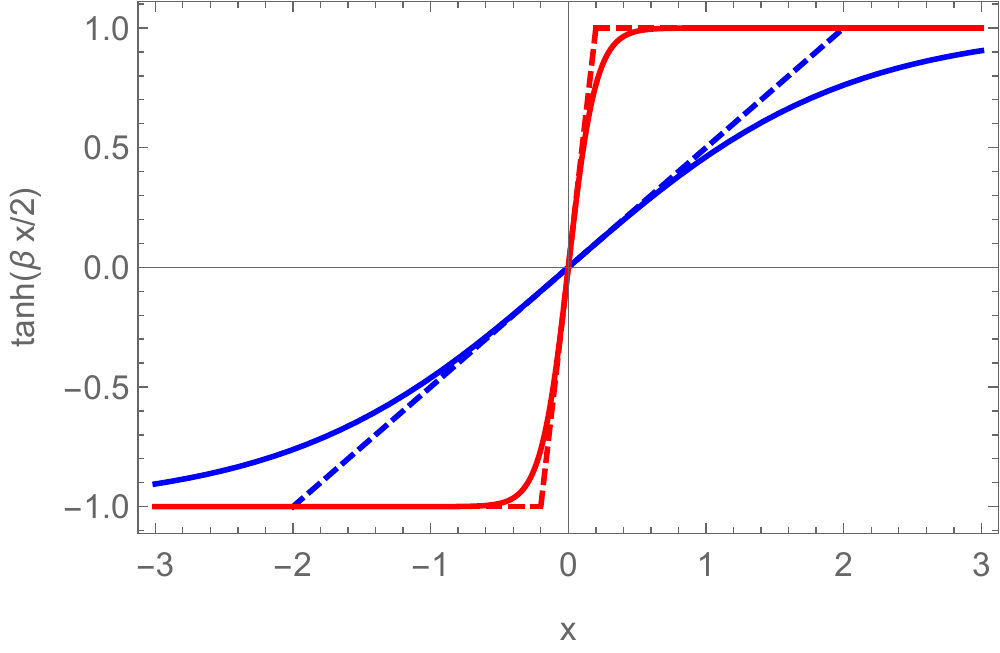}\hspace{10mm}\includegraphics[width=7.3cm]{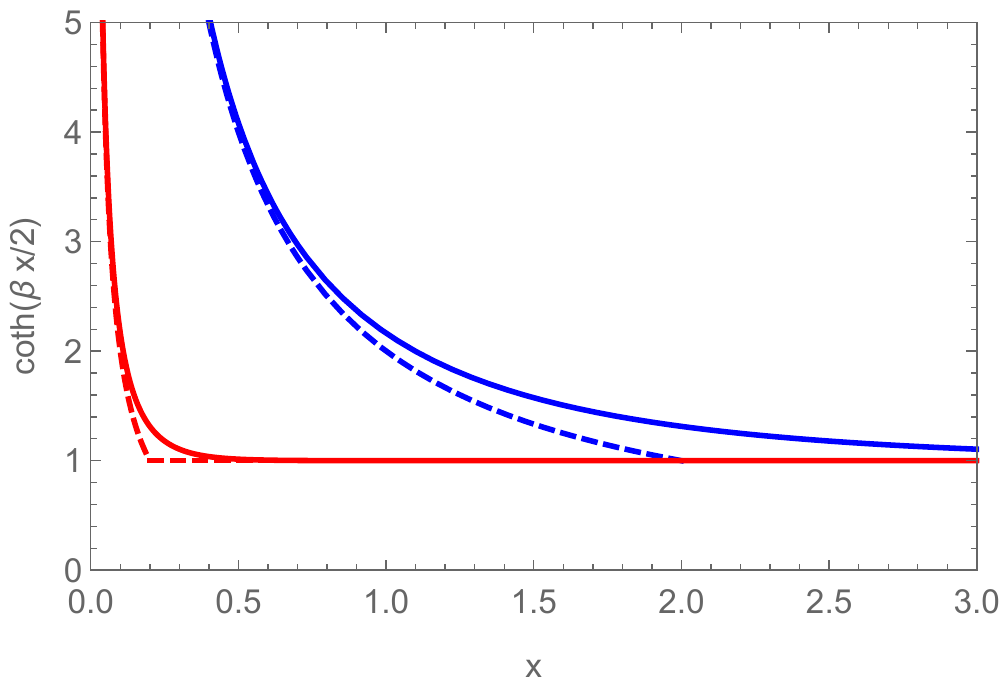}
\caption{Exact (solid line) and approximate (dashed line) representations of $\tanh(\beta x/2)$ from Eq.~\eqref{eq:FermiD}, left panel, $\coth(\beta x/2)$  from Eq.~\eqref{eq:BoseD}, right panel, for two temperatures $k_{B}T= \Delta, 0.1 \Delta$, blue and red lines, respectively.  The algebraic representation fits the Fermi and Bose distributions quite well for all temperatures, but it is quite precise for low and high ones.    \label{fig:tanh} }
\end{figure*}

We can use interpolation formulas with linear fractions  replacing the spectral functions in the regions of low and high frequencies.  We use the following approximation 
\begin{subequations}\label{eq:Distributions}
\begin{multline}\label{eq:FermiD}
\int_{-\infty}^{\infty} dx f(x) F(x) = \int_{0}^{\infty} dx \left[F_{+}(x) \phantom{\frac 12}
\right. \\ \left.
-\ \tanh\left(\frac{\beta x} 2\right)F_{-}(x)\right] \to \int_{0}^{\infty} dx F_{+}(x)
\\
 -\ \frac \beta{2}\int_{0}^{2/\beta} dx x F_{-}(x) -  \int_{2/\beta}^{\infty}dx F_{-}(x)  \,
\end{multline}
for the Fermi integral and 
\begin{multline}\label{eq:BoseD}
\int_{-\infty}^{\infty} dx b(x) F(x) = - \int_{0}^{\infty} dx \left[F_{+}(x) \phantom{\frac 12}
\right. \\ \left.
 - \coth\left(\frac{\beta x}2\right) F_{-}(x)\right] \to - \int_{0}^{\infty} dxF_{+}(x)  
 \\
 +\ \frac 2{\beta}\int_{0}^{2/\beta} \frac{dx}x F_{-}(x) +  \int_{2/\beta}^{\infty}dx F_{-}(x)   \,
\end{multline}
\end{subequations}
for the Bose integrals. We introduced symmetric and antisymmetric functions, 
$F_{+}(x) = \frac 12\left[F(x) + F(-x) \right]$ and
$F_{-}(x) = \frac 12\left[F(x) - F(-x) \right]$.
These approximate formulas well reproduce the integrals with the Fermi and Bose distribution functions from low to high temperatures, see Fig.~\ref{fig:tanh}.   

The explicit approximate analytic expressions for the integrals in Eqs.~\eqref{eq:g-full}-\eqref{eq:Dprime} are
\begin{subequations}\label{eq:gA}
\begin{align}\label{eq:g-low}
g &\doteq  -\frac{\beta}{2\pi}\arctan\left(\frac 2{\beta\Delta} \right) \,,
\\
\label{eq:DR-Lambda}
 A  &\doteq \frac{1 - a}{4\pi|g|\Delta} \left[\beta \Delta \arctan\left( \frac 2{\beta\Delta }\right) + \frac{2\beta^{2}\Delta^{2} }{4 + \beta^{2}\Delta^{2}} \right]  \,,
\end{align} 
\end{subequations}
\begin{subequations}\label{eq:y-lowT}
\begin{multline}\label{eq:y0-lowT}
y_{0} = \frac{1}{2\pi|g| \left( A - a\right)^{2}}\left\{\frac A\Delta\ln \frac{A^{2}\left(4 + \beta^{2}\Delta^{2}\right)}{4A^{2} + \beta^{2}a^{2}\Delta^{2}} 
\right. \\ \left.
+\ \beta a \left[ \arctan\left( \frac 2{\beta\Delta}\right)  - \arctan\left(\frac {2A}{\beta a\Delta}\right)\right]\right\}
\end{multline} 
and
\begin{multline}\label{eq:Deltay-lowT}
\Delta y = \frac{2A}{\pi|g|\Delta(A^{2} - a^{2})}
\\
 \times\left[\frac{A^{2} + a^{2}}{A^{2} - a^{2}} \frac{A}{a\beta \Delta}\arctan\left(\frac{A}{a\beta \Delta} \right) 
\right. \\\left.
 - \frac{2A^{2}}{A^{2} - a^{2}}\frac 1{\beta \Delta} \arctan\left( \frac 1{\beta\Delta}\right) 
-\ \frac 1{1 + \beta^{2}\Delta^{2}}\right] \,.
\end{multline}
\end{subequations}
The above equations are used to determine the dimensionless Kondo scale by combining Eqs.~\eqref{eq:bara-explicit} to a single equation
\begin{multline}\label{eq:a-SC}
\frac{(1 - a)^{3}}4\left[1 + \sqrt{1 + \frac 4{3y(1 - a)^{2}}} \right]^{3}
\\
= \frac{u}{y}\left[1 + \sqrt{1 + \frac 4{27}\frac 1{u^{2}y}} \right]\,
\end{multline}
with $u= U|g|$.

\begin{figure}
\includegraphics[width=7.5cm]{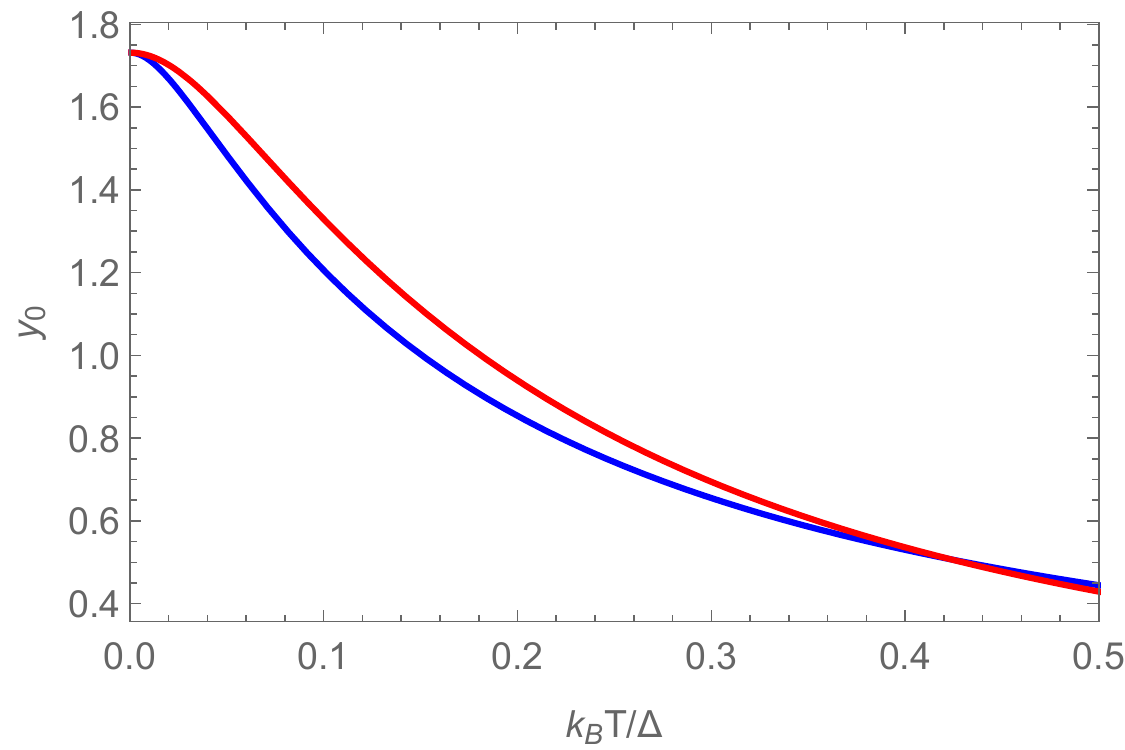}\hspace{10mm}\includegraphics[width=7.5cm]{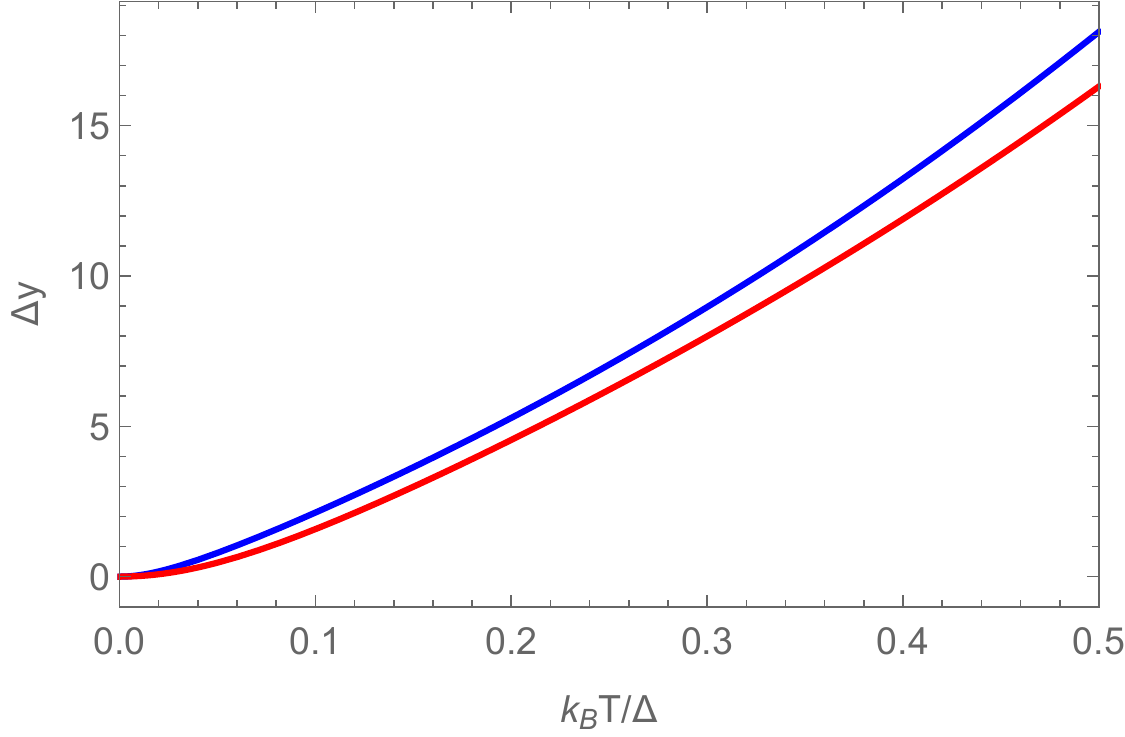}
\caption{Comparison of the exact, Eqs.~\eqref{eq:X} (blue line), and approximate, Eqs.~\eqref{eq:y-lowT} (red line), representations of $X_{0}$, left panel, and $\Delta X$, right panel, for fixed parameters $A =1$ and $a=0.1$. The precision of the algebraic approximation of the trigonometric functions increases with decreasing  Kondo scale $a$.    \label{fig:X-diff} }
\end{figure}

The algebraic approximation for the Fermi and Bose distribution functions used to reach the above  analytic expressions proves to be quite accurate, in particular in the Kondo limit $a\ll 1$ as shown in Fig.~\ref{fig:X-diff} for integrals $y_{0}$ and $\Delta y$.  

We are interested in the  low-temperature limit,  $\beta\Delta\to \infty$, in the strong-coupling regime, $a\to 0$. It follows from the above expressions that the parameter deciding how this limit looks like is  $\beta a\Delta$. We find two asymptotic regimes,  $\beta a\Delta\to \infty$ and $\beta a\Delta\to 0$ at low temperatures.  The former limit leads to the Fermi liquid, while the latter to a magnetic criticality with the Curie-Weiss susceptibility. The Fermi liquid is recovered for $\Delta y \ll y_{0}$ while the Curie-Weiss  for the opposite limit, $\Delta y \gg y_{0}$. 

All the two-particle parameters used in this approximation, $A$, and $y$, are functions of the Kondo scale $a$ determined at the end from Eq.~\eqref{eq:bara-explicit}.  Its temperature dependence is decisive for the determination of the Curie-Weiss behavior.  Both the temperature and the Kondo scale must be small. The expressions for the integrals $y_{0}$ and $\Delta y$  reduce in the Kondo limit $a\to 0$ and in the leading order of $T\to 0$ to         
\begin{subequations}\label{eq:y-Kondo}
\begin{align}\label{eq:Deltay-Kondo}
\Delta y &\doteq \frac 2{a \beta^{2}\Delta^{2}} \frac{\arctan\left(\displaystyle{\frac{A}{a\beta \Delta}}\right)}{\arctan\left(\displaystyle{\frac 2{\beta\Delta}} \right)}\,,
\\\label{eq:y_0-Kondo}
y_{0} &\doteq \frac 2{A\beta \Delta} \frac{\ln \left(\displaystyle{\frac {A}a}\right)}{\arctan\left(\displaystyle{\frac 2{\beta\Delta}}\right)} \,.
\end{align}
\end{subequations}
They replace Eqs.~\eqref{eq:y-lowT} in the asymptotic Kondo limit. 

The logarithm on the right-hand side of Eq.~\eqref{eq:y_0-Kondo} is responsible for suppressing the magnetic transition in the SIAM at zero temperature, the emergence of the exponential Kondo scale in the magnetic susceptibility, and the narrow central quasiparticle peak in the spectral function in the strong-coupling regime.

\subsection{Low-temperature regime}

  It is evident from Eq.~\eqref{eq:Deltay-Kondo} that  the Fermi-liquid regime corresponds to the low-temperature limit $a\beta\Delta  \gg A$  with $A \approx 1 $ for $\beta\Delta\to \infty$. We start with the  zero-temperature solution that reduces in the analytic low-frequency approximation to  
\begin{subequations}\begin{align}
g_{0} & = - \frac 1{\pi\Delta} \,,\\
A_{0} &= 1 - a_{0}\,, \\
y_{0} &= \frac{1}{1 - 2a_{0}}\left[\frac{1 - a_{0}}{1 - 2a_{0}} \ln\left(\frac 1{a_{0}}  - 1\right) - 1 \right] \,,
\end{align} \end{subequations}
with the Kondo scale determined from Eq.~\eqref{eq:a-SC}. The subscript $0$ refers to the values at zero temperature. Solving these equations in the strong-coupling limit $U \to \infty$ we recover the Kondo scale $a_{0} = \exp(-U/\pi \Delta)$. The exact Bethe-ansatz scale for the Lorentzian density of states is $a=\sqrt{U/2\Delta}\exp(-\pi U/8\Delta)$.\cite{Tsvelick:1983aa} Although the non-universal exponential prefactor (depending on the density of states) $\pi^{2}/8$ and the logarithmic correction $\ln\sqrt{􏰘U/2\Delta}$ are not reproduced, the universal linear dependence of the exponent of the Kondo scale on the interaction strength U is maintained.

The leading temperature contribution is quadratic, $\delta_{T} = 1/\beta^{2}\Delta^{2}$. We first evaluate the explicit temperature dependence for the fixed self-consistent parameters, $g, X= X_{0} +\Delta X, A, a$, and  $\Lambda$. We obtain
\begin{subequations}\label{eq:T2correction}
\begin{align}
\delta_{T} g &= \frac{4}{3\pi} \frac 1{\beta^{2}\Delta^{3}} \,,\\
\delta_{T} A &= - \frac{8\Lambda_{0}}{3\pi \Delta} \frac 1{\beta^{2}\Delta^{2}} \,,\\
\delta_{T} X &= - \frac {4\left(\Lambda_{0} - \pi a\Delta\right)}{3\pi^{2} \Delta^{2}}\frac{1}{\beta^{2}\Delta^{2}a^{2}} \,,
\end{align} \end{subequations}
with $\Lambda_{0} = (1 - a_{0})\pi\Delta$.

To add the temperature dependence  of the self-consistent parameters we need to evaluate partial derivatives, all taken at $T =0$. We use the derivatives with respect to the Kondo scale $a$  and the effective interaction $\Lambda$. Parameter $A$ can be determined explicitly and does not enter the self-consistency at low-temperatures. The derivatives of the $X$ integral are   
\begin{subequations}\begin{multline}
\frac{\delta X_{0}}{\delta a_{0}} = \frac {\pi\Delta}{ (\Lambda_{0} - \pi a\Delta)^{3}}
\\
\times \left[2 \Lambda_{0} \ln\left(\frac{\Lambda_{0}}{\pi a \Delta}\right) - \frac{\left(\Lambda_{0} - \pi a\Delta\right)^{2}}{\pi a\Delta}\right] \,,
\end{multline}
\begin{multline}
\frac{\delta X_{0}}{\delta \Lambda_{0}} = - \frac {1}{ (\Lambda_{0} - \pi a\Delta)^{3}}
\\
\times \left[(\Lambda_{0} + \pi a\Delta) \ln\left(\frac{\Lambda_{0}}{\pi a\Delta} \right) - 2(\Lambda_{0} - \pi a\Delta)\right] \,.
\end{multline}\end{subequations}
We will also need the derivatives of the right-hand side of Eq.(7b)  with respect to parameters $a$ and $X$,
\begin{subequations}\begin{align}
\frac{\delta\Lambda}{\delta X} &= - \frac{1}{X_{0}} \frac{U - \Lambda_{0}}{\sqrt{1 + 4(1 - a_{0})UX_{0}}} \,,\\
\frac{\delta\Lambda}{\delta a} &= - \frac{X_{0}}{1 - a_{0}} \frac{\delta\Lambda}{\delta X}\,.
\end{align}\end{subequations}

The total temperature variations of $A$ and the Kondo scale $a$ are fully determined by the non-self-consistent temperature dependence from Eqs.~\eqref{eq:T2correction} and the variation of the effective interaction $\delta\Lambda$ 
\begin{align}
\delta A &= - \frac{8\Lambda_{0}}{3\pi \Delta} \frac 1{\beta^{2}\Delta^{2}} + \frac 1{\pi\Delta}\delta\Lambda \,,\\
\delta a &= \Lambda_{0}\delta_{T} g + g_{0}\delta\Lambda\,. 
\end{align} 
We determine the variation of the effective interaction by putting together the above partial derivatives. The result is  
\begin{multline}
\left\{1 - \frac{\delta\Lambda}{\delta X}\left[\frac{\delta X_{0}}{\delta\Lambda_{0}} - \frac 1{\pi\Delta}  \left(\frac{\delta X_{0}}{\delta a_{0}} - \frac{X_{0}}{1 - a_{0}}\right)\right]\right\}\delta\Lambda 
\\
= \frac{\delta\Lambda}{\delta X}\left[\delta_{T}X + \Lambda_{0}\left(\frac{\delta X_{0}}{\delta a_{0}} - \frac{X_{0}}{1 - a_{0}}\right) \delta_{T}g \right]\,.
\end{multline}
The temperature variation of the effective interaction is positive, that is, it increases with temperature and the renormalization of the bare interaction decreases. In the Kondo strong-coupling limit $U\to\infty$ with $a\to 0$ we obtain
\be
\delta\Lambda = \frac{4\pi}{3\Delta} e^{U/\pi\Delta}\ k_{B}^{2} T^{2} \,.
\ee  
The effective interaction is, however, not a strictly monotonic function of temperature. Its slope changes around the Kondo temperature at which the thermal fluctuations start to dominate and the low-temperature Fermi-liquid behavior goes over into the Curie-Weiss regime.\cite{Janis:2020aa}

\subsection{Curie-Weiss regime}

The dominant thermal fluctuations contributing to the renormalization of the bare interaction are due to integral $\Delta X$. They start to control the approximation above the Kondo temperature at which $X_{0} = \Delta X$. The  magnetic susceptibility may take the form of the Curie-Weiss law only if $\beta a\Delta \ll 1$ and simultaneously $\beta\Delta\gg 1$.  If we use a dimensionless parameter 
\begin{equation}
\alpha = \frac 1{2}\left[\frac{\beta\Delta}{2}\arctan\left( \frac 2{\beta\Delta}\right) + \frac{\beta^{2,.}\Delta^{2}}{4 + \beta^{2}\Delta^{2}} \right]
\end{equation}
then $A= \Lambda \alpha/\pi\Delta$ and the $X$ integrals in this limit are
\begin{subequations}
\begin{align}
X_{0} &= \frac1{\Lambda\alpha} \ln\left(\frac{\Lambda\alpha}{\pi a\Delta}\right) \,, \\
\Delta X &= \frac  2{\pi\beta a \Delta^{2}}\arctan\left(\frac{\Lambda \alpha}{\pi\beta a\Delta^{2}}\right) \,.
\end{align}\end{subequations}
Assuming further that
\begin{equation}\label{eq:Lambda-alpha}
\frac{\alpha}{\pi} \frac{\Lambda}{\Delta} \gg a\beta \Delta
\end{equation}
we obtain 
\begin{equation}\label{eq:X-thermal}
X = X_{0} + \Delta X = \frac 1{\Lambda \alpha} \ln\left(\frac{\Lambda \alpha}{\pi a\Delta}\right) + \frac 1{a\beta \Delta^{2}} \,.
\end{equation}

The equations for the effective interaction and the Kondo scale $a\ll 1$ reduce to
\begin{align}\label{eq:Lambda-thermal}
\Lambda &= \sqrt{\frac{U}{ X}}  \,, \\ \label{eq:a-thermal}
1 &= \frac{U g^{2}}X \,.
\end{align}
We determine $X$ and $\Lambda$ from the above equations and are left with a single equation for the Kondo scale
\begin{equation}
Ug^{2} =  \frac {|g|}{\alpha} \ln\left(\frac{\alpha}{a|g|\Delta}\right) + \frac 1{\beta a \Delta^{2}} \,.
\end{equation}
The second term on the right-hand side of this equation should dominate the first one to  reach the Curie-Weiss regime. That is,
\begin{equation}\label{eq:T-upper}
\frac {|g|}{\alpha} \ln\left(\frac{\alpha}{a|g|\Delta}\right) \ll \frac 1{\beta a \Delta^{2}} = Ug^{2} \,. 
\end{equation}
The solution for the small Kondo scale in the Curie-Weiss regime then is
\begin{equation}\label{eq:a-CW}
a = \frac{k_{B}T}{Ug^{2}\Delta^{2}}\,.
\end{equation}

The conditions to be fulfilled to reach the Curie-Weiss regime are Eqs.~\eqref{eq:Lambda-alpha} and~\eqref{eq:T-upper}. Inserting the solution for the Kondo scale into Eq.~\eqref{eq:Lambda-alpha} we obtain an upper order-of-magnitude bound on temperature
\begin{equation}\label{eq:T-upper2}
  k_{B}T \ll \frac 1\pi U\alpha |g|\,.
\end{equation}
We obtain the lower temperature bound for the Curie-Weiss regime  from Eq.~\eqref{eq:T-upper} by using the solution for the Kondo scale from Eq.~\eqref{eq:a-CW}
\begin{equation}
k_{B}T \gg \frac 1\pi U\alpha |g|\Delta e^{-U|g|\alpha} \,. 
\end{equation}  

\begin{figure}
\includegraphics[width=7.5cm]{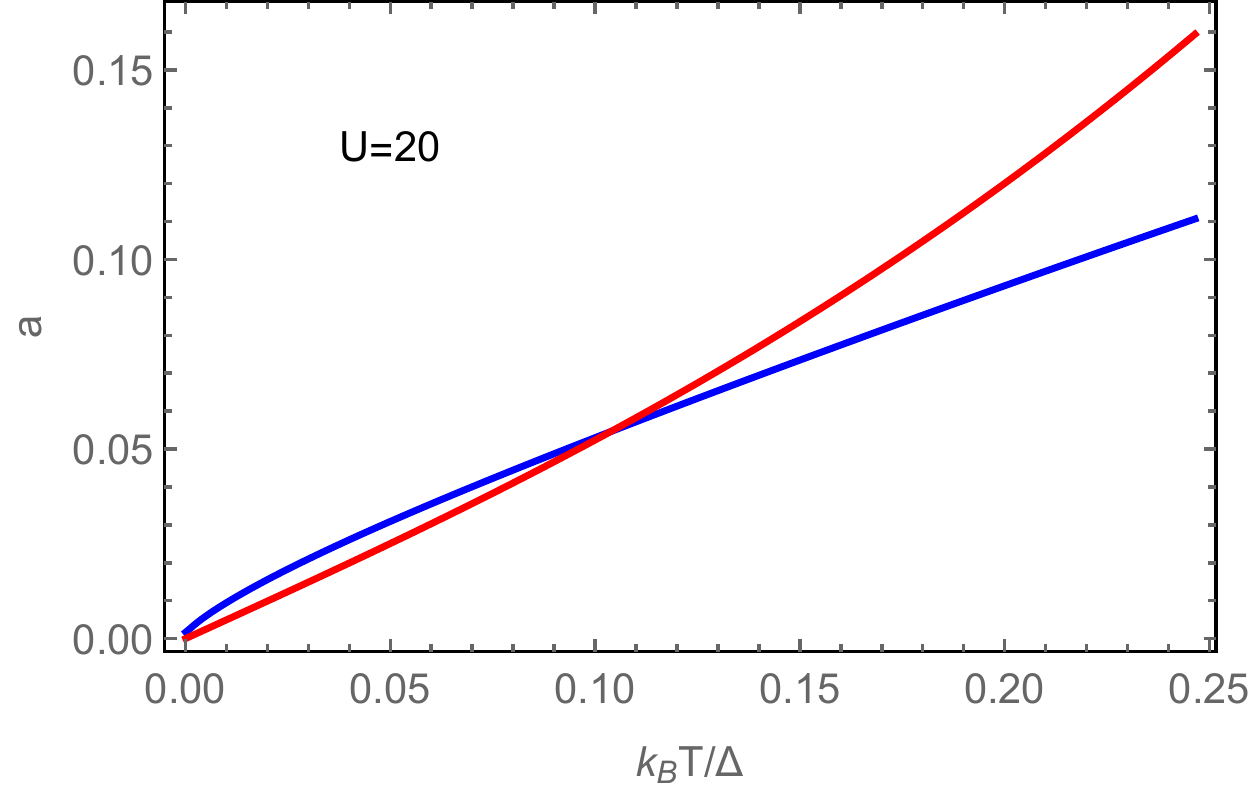}\hspace{10mm}\includegraphics[width=7.3cm]{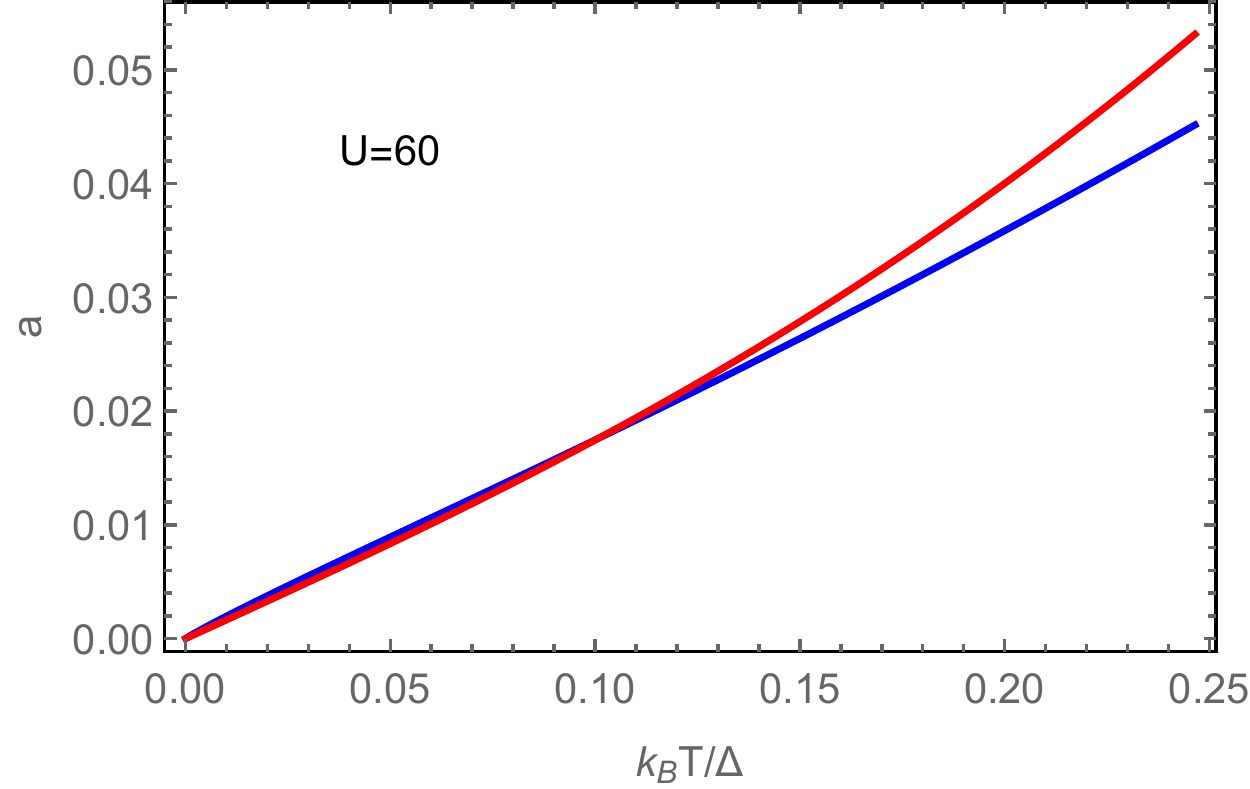}
\caption{The Kondo dimensionless scale $a$ calculated from Eq.~\eqref{eq:a-def} blue line, and from the asymptotic formula in the Curie-Weiss regime, Eq.~\eqref{eq:a-CW}, red line, for two values of the interaction strength. Linear temperature dependence sets in only on a small temperature interval and for strong electron repulsion.    \label{fig:a-comp} }
\end{figure}

\begin{figure}
\includegraphics[width=7cm]{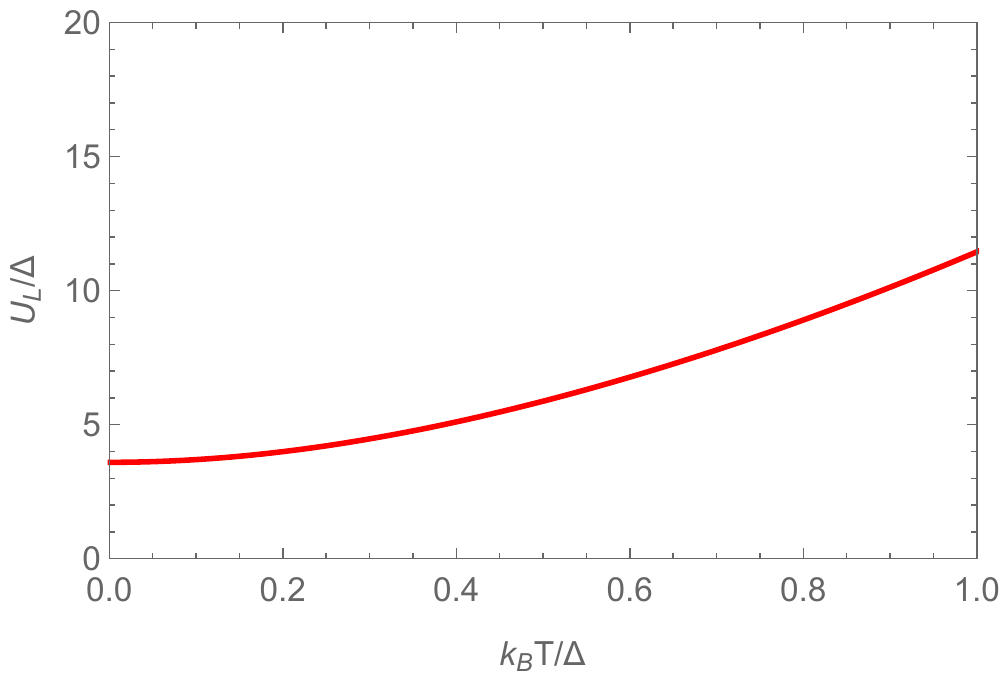}
\caption{ The lower bound on the interaction strength above which we can expect the Curie-Weiss magnetic susceptibility as defined on the left-hand side of Eq.~\eqref{eq:U-lower}. The bound increases fast with the increasing temperature.  \label{fig:UL} }
\end{figure}

Next, we have to satisfy the condition for criticality in the Kondo regime, $a\to 0$. The Kondo scale from quantum (zero-temperature) fluctuations and integral $X_{0}$ should be very small
\be
a > a_{Q} = \frac{\alpha}{|g|\Delta} e^{-U|g|\alpha} \ll 1\,.
\ee
 It sets the lower bound for the Kondo scale at non-zero temperatures. It is small only in strong-coupling when the exponent on the right-hand side is sufficiently big, namely   
\begin{equation}\label{eq:U-lower}
U \gg  U_{L} = \frac{1}{ |g| \alpha} \ln\left(\frac{\alpha}{|g|\Delta} \right) \,.
\end{equation}
The boundaries for the linear temperature dependence of the Kondo scale $a$ restrict the Curie-Weiss region to a rather narrow interval and for very strong interactions as shown in Fig.~\ref{fig:a-comp}.   %
The lower bound $U_{L}$ on the interaction strength increases significantly with increasing temperature as shown in Fig.~\ref{fig:UL}.  

The magnetic susceptibility is directly connected to the Kondo scale $a$, Eq.~\eqref{eq:chiinv-full},
\begin{equation}\label{eq:CW-susceptibility}
 \frac{\chi^{T}}{\mu_{0}\mu_{B}^{2}} = \frac{2|g|}a = \frac{2|g|^{3}\Delta^{2}}{k_{B}T}U\,.
\end{equation}
 Combining this representation with the solution for the Kondo scale from Eq.~\eqref{eq:a-CW} we obtain an equation for the effective Curie constant
\begin{equation}\label{eq:C-Constant}
C = \frac{U|g|^{3}\Delta^{2}}{k_{B}} \,.
\end{equation} 
The Curie-Weiss law becomes pronounced if the Curie constant is only weakly dependent on temperature. It is  when 
\begin{multline}
\frac T{C}\frac{d C}{d T}= \frac {2T}g \frac{d g}{d T} = \frac 1{\displaystyle{\arctan\left(\frac 2{\beta\Delta}\right)}} 
\\
\times\left[\arctan\left(\frac 2{\beta\Delta}\right) - \frac{2\beta\Delta}{\beta^{2}\Delta^{2} + 4} \right] \ll  1 \,.
\end{multline}
It leads to an upper bound on temperature that at low temperatures is  $k_{B} T \ll \Delta\sqrt{3/{8}}$ below which we can observe the Curie-Weiss behavior.. It is a stronger upper bound on the validity of the Curie-Weiss susceptibility than that from Eq.~\eqref{eq:T-upper2}.
 
The full estimate for the low-temperature behavior of the magnetic susceptibility above the Kondo temperature and in the region of the Curie-Weiss linear response is 
\begin{equation}\label{eq:chi-CW}
\frac{\chi^{T}}{\mu_{0}\mu_{B}^{2}} =  \frac{\displaystyle{\frac{U\beta^{3}\Delta^{2}}{4\pi^{3}k_{B}}\arctan^{3}\left(\frac 2{\beta\Delta}\right)}}{T + \displaystyle{\frac{\beta^{2}\Delta^{2}}{4\pi^{2}k_{B}}\arctan^{2}\left(\frac 2{\beta\Delta}\right)U e^{-U/\pi \Delta}}} \,,
\end{equation}
where we shifted the origin of the linear temperature dependence to the Kondo temperature to get a better fit for the behavior close to the Kondo temperature.  

The lower bound for the validity of the Curie-Weiss law is the Kondo temperature $T_{K}$ defined from equality of quantum and thermal fluctuations expressed by  an equation $\Delta X = X_{0}$.\cite{Janis:2020aa}  The magnetic susceptibility is then well approximated by the Curie-Weiss law if  $\Delta X \gg X_{0}$, which sets the lower temperature bound. The order-of-magnitude temperature bounds on the validity of the Curie-Weiss law in the SIAM are
\begin{equation}\label{eq:T-CW}
\sqrt{\frac{3}{2}}\frac {\Delta}2 \gg k_{B}T \gg \frac U{\pi^{2}} e^{-U/\pi \Delta} \,.
\end{equation} 
\begin{figure}
\includegraphics[width=7.5cm]{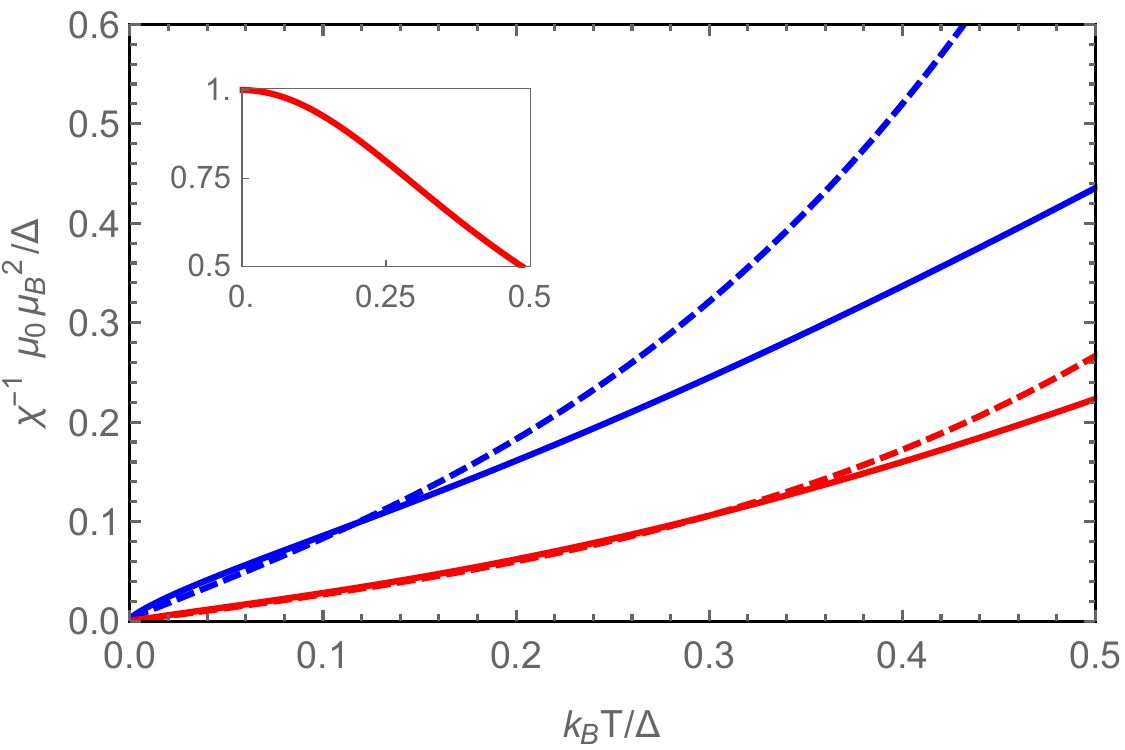}
\caption{The inverse magnetic susceptibility calculated from the exact formula, Eq.~\eqref{eq:chiinv-full} (solid lines), and from the asymptotic form, Eq.~\eqref{eq:chi-CW} (dashed lines), for two values of the interaction strength, $U/\Delta=20,60$, blue and red lines, respectively. The inset shows the temperature dependence of the ratio of the Curie constants $C/C_{0}$. \label{fig:chi-inv}}
\end{figure}

The Curie-Weiss susceptibility exists only in strongly correlated systems and sufficiently above the Kondo temperature but still at low temperatures, small fractions of the bandwidth, as demonstrated in Fig.~\ref{fig:chi-inv}. We can see a very good agreement of the asymptotic expression from Eq.~\eqref{eq:chi-CW} and the full numerical solution  for interactions $U>20\Delta$.  Although we may observe almost linear temperature dependence of the inverse susceptibility also at rather high temperatures, its origin there is no longer in the critical behavior of the magnetic transition at which the Kondo scale vanishes, $a=0$.

\subsection{High-temperature regime} 

The Curie-Weiss regime realizes on a temperature interval sufficiently above the Kondo temperature and sufficiently below the temperature corresponding to the band width. We know that for any finite interaction strength $U$ the high temperature limit of the magnetic susceptibility is in our units $\chi \doteq \beta/2$ [\onlinecite{Haldane:1978aa}]. This high-temperature asymptotics does not correspond to the Curie-Weiss susceptibility in  Eq.~\eqref{eq:chiinv-full} that has the high-temperature asymptotics $\beta\to 0$
\begin{equation} \label{eq:chi-CW-HT}
\frac{\chi^{T}}{\mu_{0}\mu_{B}^{2}}\doteq  \frac{\beta^{4}}{32}U \Delta^{2} \,. 
\end{equation}  
It means that there is a crossover from the Curie-Weiss linear dependence of the inverse susceptibility to the high-temperature linear dependence.  The crossover depends on the interaction strength. Equation~\eqref{eq:chi-CW-HT} holds only in the Kondo regime with $a\ll 1$. It poses a restriction on the interaction strength that for high temperature ($\beta\to 0$) from Eq.~\eqref{eq:a-CW} is  
\be
U \gg \frac {16}{\Delta^{2}} (k_{B}T)^{3}\,.
\ee
The crossover temperature from intermediate to high-temperature regimes is
\be
T_{H} = \frac 1{2k_{B}}\sqrt[3]{\frac{U\Delta^{2}}2} \,.
\ee 
The Curie-Weiss regime breaks down before this crossover temperature is reached. 

The high-temperature asymptotics of the magnetic susceptibility in the strong-coupling regime can now be assessed by using the crossover temperature $T_{H}$
\begin{subequations}
\begin{align}
\frac{\chi^{T}}{\mu_{0}\mu_{B}^{2}} &= \frac 1{2k_{B}T} \left(\frac{T_{H}}{T} \right)^{3} \qquad \mathrm{for}\quad T< T_{H} \,, \\
\frac{\chi^{T}}{\mu_{0}\mu_{B}^{2}} &= \frac 1{2k_{B}T}\phantom{\left(\frac{T_{H}}{T} \right)^{3}} \qquad \mathrm{for}\quad T> T_{H} \,.
\end{align}
\end{subequations}
The crossover temperature has a similar effect as the Kondo temperature on the opposite temperature scale. The high-temperature asymptotics of the susceptibility decreases as $T^{-4}$ up to the crossover temperature $T_{H}$ above which it decreases as $T^{-1}$.

\section{Reliability and  quantitative accuracy of the approximate solution of the SIAM} 

The three temperature regimes in the SIAM were derived within the reduced two-channel parquet equations in which we neglected noncritical fluctuations in the strong-coupling Kondo regime. This regime is determined by the asymptotically vanishing dimensionless scale $a=1 + \Lambda\phi(0)$. It means, that the approximation we used is justified and qualitatively reliable in the critical region $a\ll 1$. This regime is reached in the SIAM only in the strong-coupling regime $U\to \infty$ and at very low temperatures $k_{B}T/\Delta\ll 1$. The parameter deciding which temperature regime sets in is $a\beta\Delta$. We set approximate bounds for each of the temperature regimes in the SIAM from the estimates of the behavior of this parameter. We used further approximations to derive the analytic formulas for the Kondo scale $a$, the effective interaction $\Lambda$ and the thermodynamic susceptibility $\chi^{T}$.

Our analytic approximation allowed us to disclose the mechanism for the emergence of the Curie-Weiss law in the magnetic susceptibility even in the SIAM where this behavior was overlooked.   It was mostly due to the fact that the advanced non-perturbative solutions at non-zero temperatures are purely numerical from which we cannot derive criteria for the existence of the Curie-Weiss law. Moreover, the Curie-Weiss regime sets in only for extremely strong interactions that were out of interest and were not studied. Nevertheless, one should test the validity of the assumptions used in the approximate Curie-Weiss behavior. 
 
\begin{figure}
\includegraphics[width=7.5cm]{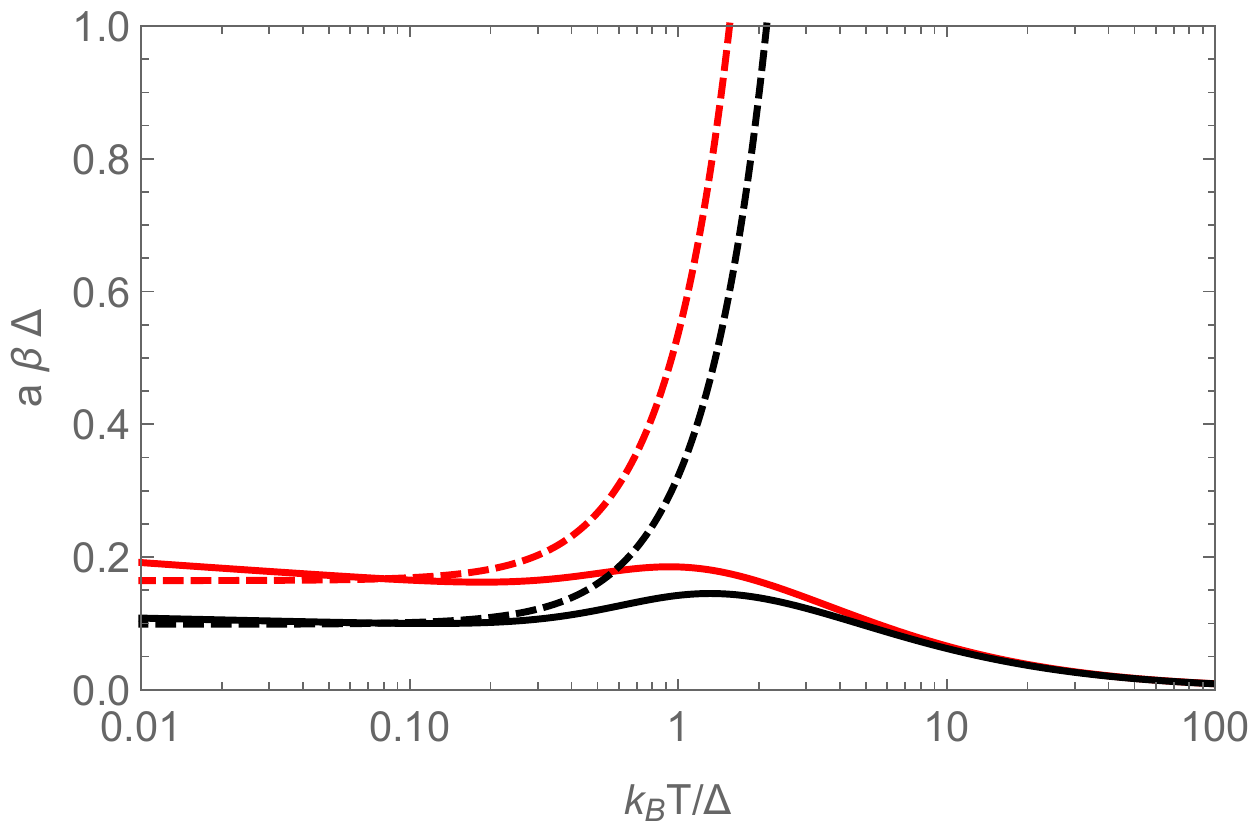}
\caption{The controlling parameter $a\beta\Delta$ decisive for setting the temperature regime. The Fermi-liquid sets for $a\beta\Delta\gg 1$, while  the Curie-Weiss for $a\beta\Delta\ll 1$. The full solution (solid lines) and the asymptotic solution from Eq.~\eqref{eq:a-CW} (dashed lines) are compared for two interaction strengths $U= 60\Delta$ (red curves) and $U= 100\Delta$ (black curves).   \label{fig:aBeta} }
\end{figure}

The Curie-Weiss regime is characterized by a linear dependence of the Kondo scale $a$ on temperature $T$. It is expressed by Eq.~\eqref{eq:a-CW}. This behavior was derived with two assumptions, $a\beta\Delta\ll 1$ and $\Delta y \gg y_{0}$. We plotted in Fig.~\ref{fig:aBeta} the temperature dependence of the controlling parameter $a\beta\Delta$ for two interaction strengths, $U= 60\Delta,100\Delta$. Both the full and the asymptotic solution coincide around $k_{B}T\approx 0.1\Delta$. The higher the interaction the smaller the parameter and the broader the interval on which the full and asymptotic solutions coincide are.   

\begin{figure}
\includegraphics[width=7.2cm]{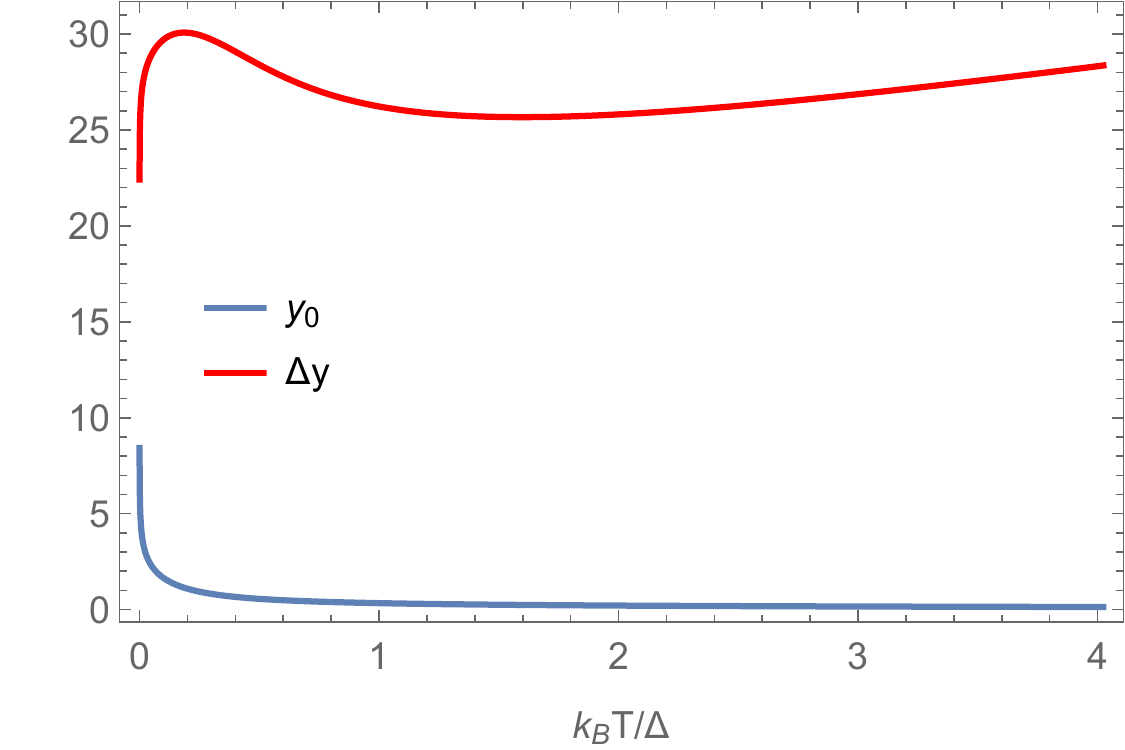}\hspace{10mm}\includegraphics[width=7.6cm]{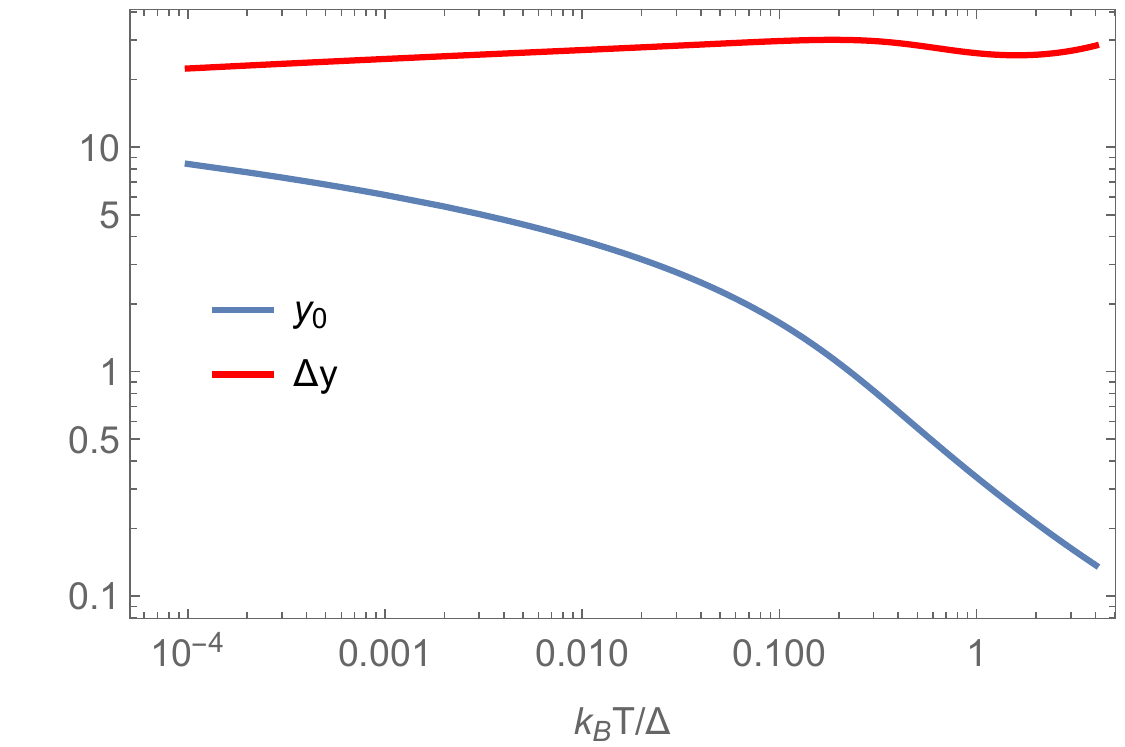}
\caption{Comparison of the two integrals $y_{0}$ and $\Delta y$ calculated from the asymptotic solution, Eq.~\eqref{eq:y-lowT} for  $U= 100\Delta$ in the linear scale, upper panel, and log-log scale, lower panel.  The Curie-Weiss regime sets in for $\Delta y \gg y_{0}$.    \label{fig:y0Dy} }
\end{figure}

The other criterion for the existence of the Curie-Weiss law is the dominance of the thermal fluctuations represented by integral $\Delta y$ over the quantum ones in integral $y_{0}$.  We plotted the two integrals in linear and log-log scales in Fig.~\ref{fig:y0Dy}. The figure confirms that $\Delta y \gg y_{0}$ around $k_{B}T\approx 0.1\Delta$. 

These test criteria are internal ones using the parameters introduced in our solution and cannot be checked independently by other approaches. We plotted a thermodynamic quantity $T\chi^{T}$ available from the exact Bethe-ansatz solution\cite{Okiji:1983aa} and the Monte-Carlo simulations\cite{Fye:1988aa} for intermediate interaction strengths. The output of our approximation is plotted in Fig.~\ref{fig:Tchi} for interactions strengths $U= 20\Delta,60\Delta,100\Delta$. The Curie-Weiss behavior is indicated by a plateau at low temperatures determining the effective Curie constant $C$ from Eq.~\eqref{eq:C-Constant}. The parameter falls down to zero below the Kondo temperature. We can see only a marginal temperature dependence of the Curie constant around $k_{B}T\approx 0.1\Delta$ where the Kondo scale depends linearly on temperature.  The Fermi-liquid regime sets in for much lower temperatures beyond the plotted region. 
\begin{figure}
\includegraphics[width=7.5cm]{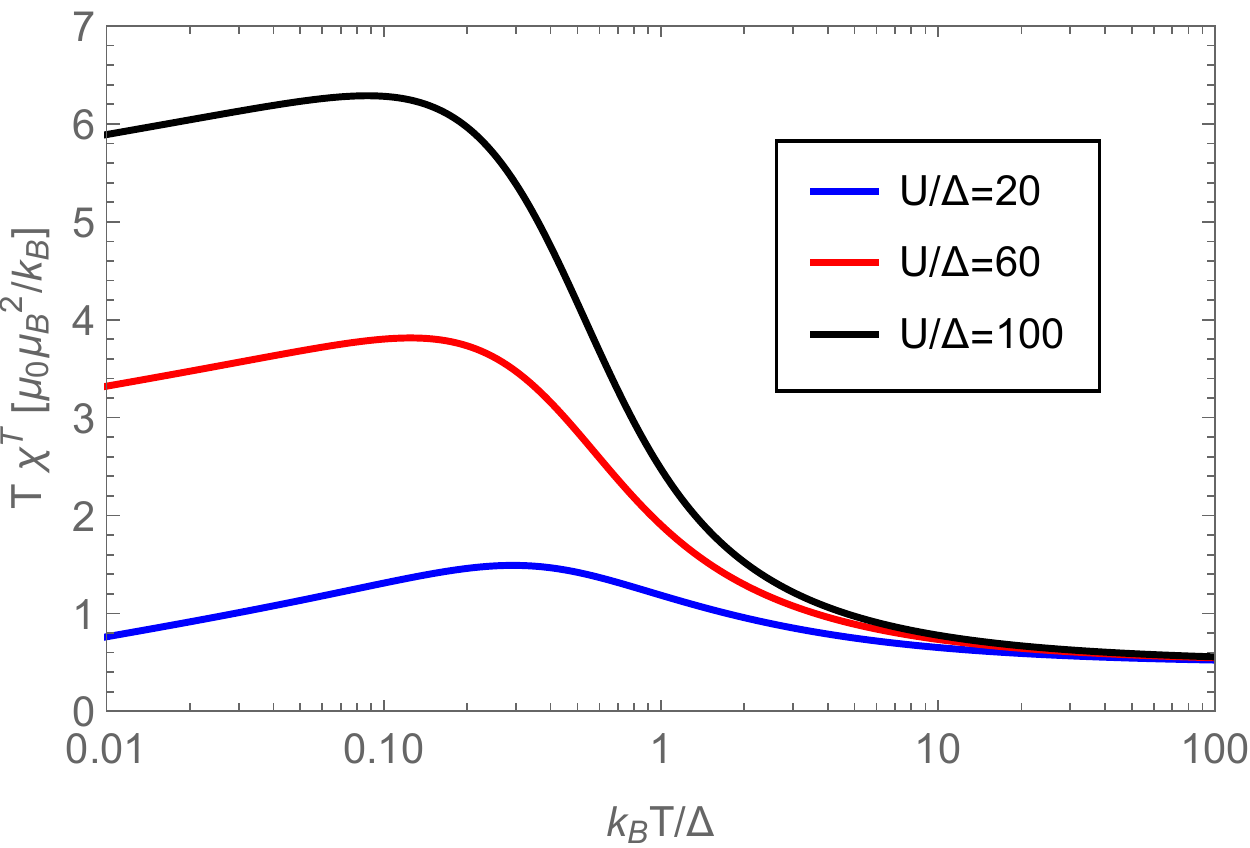}
\caption{Thermodynamic parameter $T\chi^{T}$ demonstrating the transition from the high-temperature to the Curie-Weiss regime for interaction strengths $U=20\Delta,60\Delta,100\Delta$, blue, red, and black lines.   \label{fig:Tchi} }
\end{figure}

Neither Bethe-ansatz nor the quantum Monte-Carlo (QMC) simulations are available for the interaction strengths of order $U\approx 50\Delta$. We hence performed our own simulations using the TRIQS/CTHYB continuous-time, hybridization expansion quantum Monte Carlo solver\cite{Seth:2016aa} to calculate the temperature dependence of the susceptibility. We used a constant tunneling density of states $\Delta(\omega)=\Theta(W^2-\omega^2)/(2W)$ with a half-bandwidth $W=100\Delta$. The results depend only weakly on the bandwidth (less than 2\% difference between $W=100\Delta$ and $W=1000\Delta$). The static susceptibility was calculated by integration of the dynamical susceptibility $\chi(\tau)=\mu_0\mu_B^2\langle m(\tau)m(0)\rangle$ where $m=n_\uparrow-n_\downarrow$. The results for $U=10\Delta,20\Delta,60\Delta$ are plotted in Fig.~\ref{fig:Tchi}. We can see that the plateau starts to form for $U>20\Delta$ and the stronger the interaction the broader and flatter the plateau is. The predicted Curie-Weiss law in the SIAM by an analytic approach  is thus  confirmed by numerical QMC simulations. The simulations indicate that our analytic estimates for the interval on which the Curie-Weiss law holds are too conservative.  Notice that the QMC simulations start showing tangible statistical errors at low temperatures and they cannot reach the Fermi-liquid regime in the strong-coupling limit.  
\begin{figure}
\includegraphics[width=8cm]{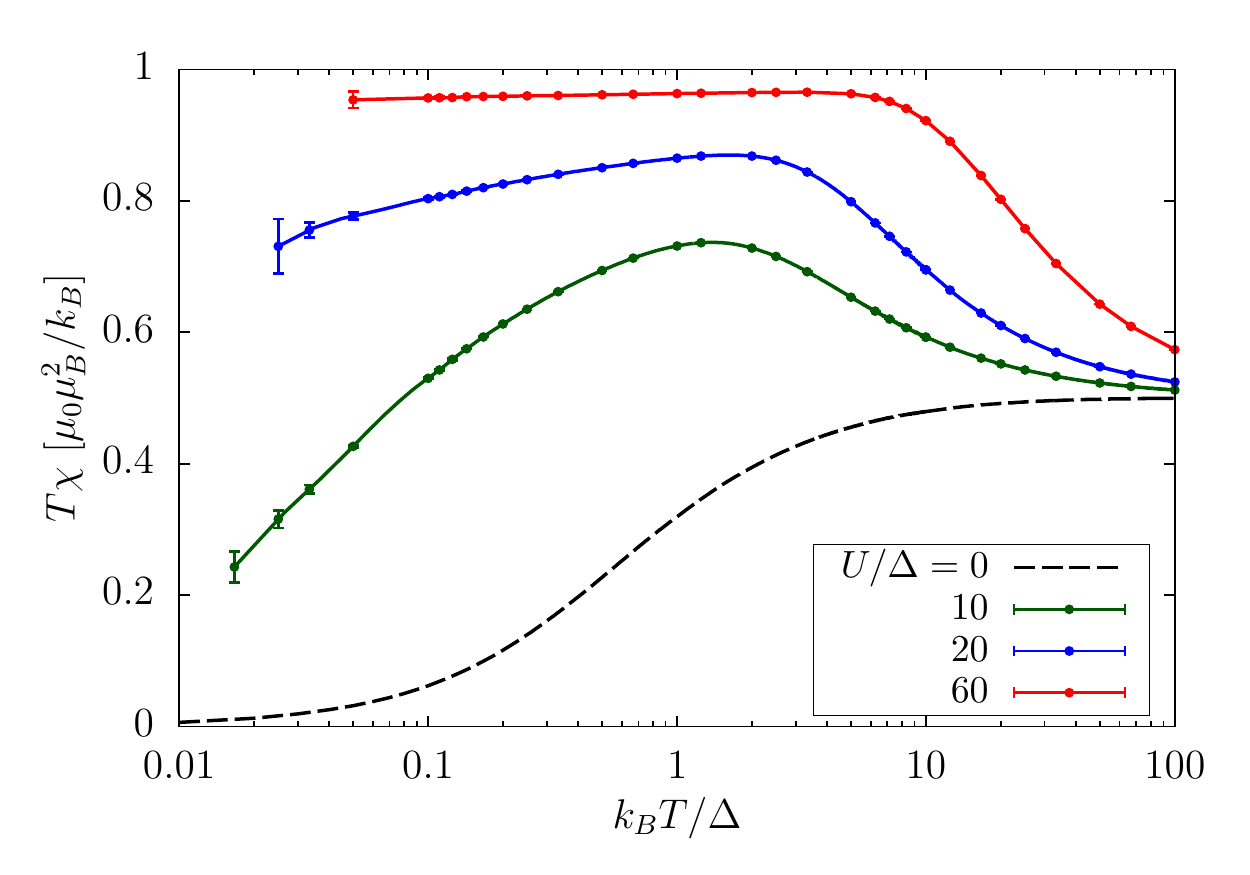}
\caption{ CT-HYB QMC data (symbols with error bars) for $T\chi$. The solid lines are a spline of the CT-HYB data and serve only as s guide for the eye. The simulations confirm the emergence of the Curie-Weiss law in the magnetic susceptibility for suﬀiciently strong electron interactions, $U > 20\Delta$ on a broader temperature interval than predicted analytically.
   \label{fig:Tchi} }
\end{figure}

The temperature dependence of $T\chi^{T}$  in the approximate analytic solution and the QMC data are qualitatively similar. The most visible difference is, however, the height of the low-temperature plateau, or the numerical value of the Curie constant. It is not so surprising, since this parameter is noncritical and does not scale with the vanishing parameter $a$. We used the bare propagators to determine the Curie constant, Eq.~\eqref{eq:C-Constant}. The bare propagators are good for determining the universal critical behavior but must be renormalized by a dynamical spin-symmetric self-energy to improve upon quantitative estimates of noncritical quantities. The dynamical or spectral self-energy is determined from the Schwinger-Dyson equation where various degrees of one-particle self-consistency can be used.\cite{Janis:2019aa} The next quantitative improvement will be achieved by considering a nontrivial dependence of the irreducible vertex $\Lambda$ on fermionic frequencies. This turns the algebraic equation determining the critical behavior integral with the necessity to diagonalize the integral kernel of the Bethe-Salpeter equation for the reducible vertex $\mathcal{K}$. The latter step goes beyond the mean-field character of the proposed analytic approximation. 

The value of the noncritical quantity $T\chi^{T}$ is then sensitive to the renormalization of the one-electron propagators we used to determine the susceptibility. If we choose a self-energy $\Sigma(\omega_{+})$ to renormalize the thermodynamic propagator and keep the irreducible vertex $\Lambda$ frequency-independent  the magnetic susceptibility will be\cite{Janis:2019aa}
\begin{multline}\label{eq:chi-renorm}
\chi = \left(2 + \Lambda \chi^{T} \right)\int_{-\infty}^{\infty} \frac{dx}{\pi}f(x)  \Im\left[G\left(x_{+} - \Sigma(x_{+})\right)^{2} \right]
\\
= - \frac 2a \int_{-\infty}^{\infty} \frac{dx}{\pi}f(x)  \Im\left[G\left(x_{+} - \Sigma(x_{+})\right)^{2} \right] \,.
\end{multline}

The dynamical self-energy is determined from the Schwinger-Dyson equation in the spin-symmetric sector not to affect the magnetic critical behavior derived with the thermodynamic propagators. Its form with the frequency-independent irreducible vertex $\Lambda$ is\cite{Janis:2019aa}   
\begin{multline}\label{eq:Sigma-SD}
\Sigma(\omega_{+}) =  U\Lambda
\\   
 \times\int_{-\infty}^{\infty}\frac{dx}{\pi } \left\{f(x + \omega) \frac{\phi( x _{-})}{ 1 + \Lambda \phi(x_{-})}  \Im G(x + \omega_{+})
\right. \\ \left.
 \ - b(x)G(x + \omega_{+}) \Im\left[\frac{\phi( x _{+})}{1 + \Lambda \phi(x_{+})} \right]  \right\}\,,
\end{multline}
\begin{figure}
\includegraphics[width=8cm]{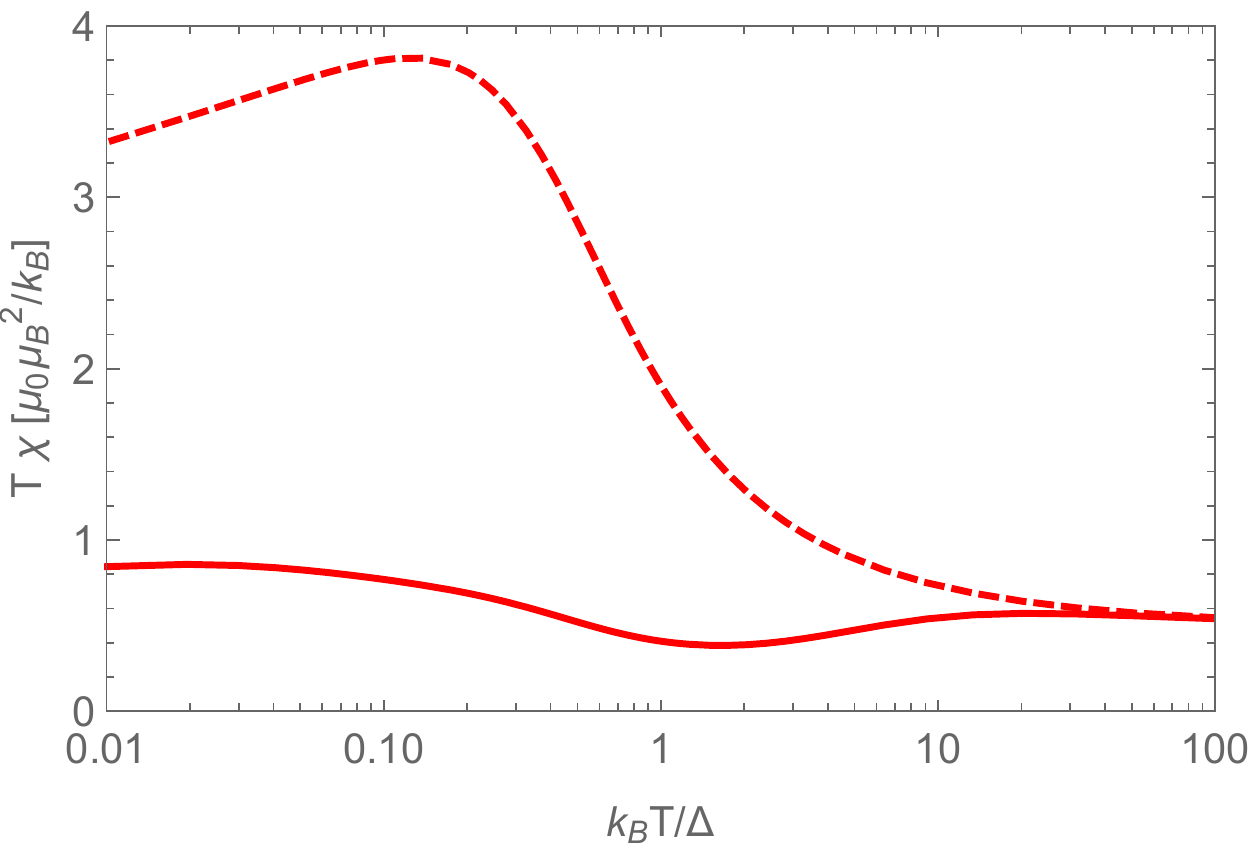}
\caption{ Thermodynamic parameter $T\chi$ calculated from Eq.~\eqref{eq:chiinv-full}, dashed line, and with the susceptibility from  Eq.~\eqref{eq:chi-renorm} renormalized by the self-energy from the lowest order in Eq.~\eqref{eq:Sigma-SD}, solid line, for $U = 60\Delta$. The value of the Curie constant was significantly decreased and became less temperature sensitive. It is closer to the value from the QMC simulations at low temperatures but misses the crossover from intermediate to high-temperature asymptotics.\label{fig:Tchi-renorm} }
\end{figure}
We used the lowest-order contribution in $\Lambda$ to this self-energy to check the dependence of parameter $T\chi$ on the dynamical corrections of the one-particle propagators. Its temperature dependence is plotted in Fig.~\ref{fig:Tchi-renorm}.   The Curie constant  is closer to the value obtained from the QMC simulations and it becomes also less temperature dependent than the one calculated from $\chi^{T}$ with the bare propagators. It misses, however, to reproduce the correlation-induced crossover downfall from intermediate to high temperatures. It means that the renormalization of the susceptibility with the lowest-order self-energy from Eq.~\eqref{eq:Sigma-SD} falls into the weak-coupling regime at intermediate temperatures before the high-temperature asymptotics sets in.  Adding more terms to the self-energy in the Schwinger-Dyson equation  decreases the value of the Curie constant even more. It is hence diﬀcult to find the appropriate renormalization of the vertex function and the self-energy to simulate quantitatively accurately  the behavior of thermodynamic quantities in the whole range of temperatures. Dynamical corrections to the irreducible vertex $\Lambda$ have to be taken into account.

\section{Extended systems}

The reduced parquet approximation can straightforwardly be extended to lattice systems. The dynamics of the two-particle functions is then determined not only by frequency but also momentum  fluctuations. That is, the dynamical variable in the two-particle integrals changes from $\omega$ to  $(\vecq,\omega)$. The Kondo scale becomes momentum dependent, $a\to a(q) = a + \Lambda Dq^{2}$, the electron-hole bubble goes over to $\phi(\omega) \to \phi(\vecq,\omega)$, and the denominator of the low-energy asymptotics of the dynamical susceptibility transforms to $1 + \Lambda\phi(\omega) \to 1 + \Lambda\phi(\vecq,\omega) \doteq a + \Lambda Dq^{2} - i\Lambda A\omega/\Delta$, where $\Delta$ is an effective bandwidth. The integral renormalizing the bare interaction goes over to
\begin{multline}
\mathcal{X} \propto \phi_{0}^{2}S_{d}\int_{0}^{2/l_{0}} \frac{d q}{(2\pi)^{d}}q^{d - 1} 
\\
\times\int_{-2t}^{2t} \frac{d\omega}{\pi}b(\omega) \Im\left[\frac 1{a + \Lambda Dq^{2} - i\pi \Lambda A\omega }  \right] \,,
\end{multline} 
where $\phi_{0}= \phi(\mathbf{0},0)$ and $S_{d}$ is the surface of the $d$-dimensional sphere. We used cutoffs for this low-energy asymptotic formula with $t$ being the hopping amplitude on the $d$-dimensional hypercubic lattice and $1/l_{0}$ is an appropriate cutoff on the momentum integration.  The momentum integral changes the low-temperature dependence and magnetic transitions with $a=0$ at non-zero temperatures may occur  only if integral $\mathcal{X}$ is finite, which happens in dimensions $d>2$. The Curie-Weiss behavior can be observed on an interval of temperatures on which the effective Curie constant  $C$ does not change much from its value at the critical point.  The Kondo scale in extended systems is inversely proportional to the spatial correlation length $\xi$, $a= \Delta^{2}/\xi^{2}$. The critical behavior of the extended systems is compatible with the Mermin-Wagner theorem.\cite{Mermin:1966aa} No long-range order exists for dimensions $d=1,2$ at non-zero temperatures,  since integral $\mathcal{X}$  is linearly and logarithmically divergent for $a=0$. This makes the reduced parquet equations  a suitable and affordable approximation for studying qualitative behavior in the  critical region of instabilities in realistic systems with strongly correlated electrons.

\section{Conclusions}

There is no transition to the magnetic state in the SIAM with $a=0$ and  that is why the Curie-Weiss law sets only for extremely strong interaction strengths, above the Kondo temperature, and beyond the Fermi-liquid regime. It is also the reason why the Curie-Weiss law has not yet been demonstrated in SIAM. The strong-coupling Kondo limit of SIAM is, however, a paradigm for the explanation of the Curie-Weiss magnetic response in metals. Our analysis has therefore a general significance beyond the impurity models.  We can draw general conclusions  if we appropriately interpret the scales used in SIAM. The Kondo scale is generalized to  $a_{c}= (T - T_{c})/T_{F}$, where $T_{c}$ is the magnetic critical and $T_{F}$ the Fermi temperature. Further on, the effective bandwidth $\Delta \sim k_{B}T_{F}$ and $A\sim \Lambda\pi \rho_{F}$ with $\rho_{F}$ being the local density of states at the Fermi energy. The critical magnetic response in metallic systems is controlled by the generalized Kondo scale $a_{c}$. The smaller the ratio $T_{c}/T_{F}$  the longer the temperature interval with the Curie-Weiss susceptibility above the critical temperature is. The Curie constant $C$ is proportional to the bare interaction $U$ while the critical temperature $T_{c}$ is proportional to the renormalized vertex $\Lambda$. The Curie-Weiss behavior in metallic systems is hence most pronounced for broad-band systems with strong and significantly screened electron interaction.

\section*{Acknowledgment}
 The research was supported by Grant No. 19-13525S of the Czech Science Foundation and INTER-COST LTC19045 (V.P.). The computational resources were supplied by the project "e-Infrastruktura CZ" (e-INFRA LM2018140) provided within the program Projects of Large Research, Development and Innovations Infrastructures.


\end{document}